%% file: main.tex
\documentclass[sigconf,dvipsnames]{acmart}

\usepackage[draft]{common}
\usepackage{etoolbox}
\usepackage{amsmath}
\usepackage{amsfonts}
\usepackage{graphicx}
\usepackage{tabularx}
\usepackage{booktabs}
\usepackage[style=base]{caption}
\usepackage{longtable}
\usepackage{multirow}
\usepackage{makecell}
\usepackage{pifont}
\usepackage[hyphens]{xurl}
\usepackage{hyperref}
\usepackage{listings}
\usepackage{tcolorbox}
\usepackage{wasysym}
\usepackage{cuted}
\usepackage{flushend}

\usepackage{tikz}
\usepackage{tikzpeople}
\usepackage{tikz-cd}
\usetikzlibrary{arrows.meta,
	positioning,
	quotes
}

\apptocmd{\sloppy}{\hbadness 10000\relax}{}{}

\setcopyright{none}
\acmDOI{}
\renewcommand\footnotetextcopyrightpermission[1]{} % removes footnote with conference information in first column

\makeatletter
\newif\ifhidelinks@hidelinks
\newcommand{\hidelinks}{%
	\hidelinks@hidelinkstrue
	\let\hidelinks@ifHy@colorlinks@status\ifHy@colorlinks
	\let\hidelinks@ifHy@ocgcolorlinks@status\ifHy@ocgcolorlinks
	\let\hidelinks@ifHy@frenchlinks@status\ifHy@frenchlinks
	\let\hidelinks@Hy@colorlink\Hy@colorlink
	\let\hidelinks@Hy@endcolorlink\Hy@endcolorlink
	\let\hidelinks@@pdfborder\@pdfborder
	\let\hidelinks@@pdfborderstyle\@pdfborderstyle
	\hypersetup{hidelinks}%
}
\newcommand{\restorelinks}{%
	\ifhidelinks@hidelinks
	\hidelinks@hidelinksfalse
	\let\ifHy@colorlinks\hidelinks@ifHy@colorlinks@status
	\let\ifHy@ocgcolorlinks\hidelinks@ifHy@ocgcolorlinks@status
	\let\ifHy@frenchlinks\hidelinks@ifHy@frenchlinks@status
	\let\Hy@colorlink\hidelinks@Hy@colorlink
	\let\Hy@endcolorlink\hidelinks@Hy@endcolorlink
	\let\@pdfborder\hidelinks@@pdfborder
	\let\@pdfborderstyle\hidelinks@@pdfborderstyle
	\fi
}
\makeatother

\begin{document}

\title{SoK: Decentralized Finance (DeFi)}

\author{Sam Werner}
\affiliation{
    \institution{Imperial College London}
    \country{}
}

\author{Daniel Perez}
\affiliation{
    \institution{Imperial College London}
    \country{}
}

\author{Lewis Gudgeon}
\affiliation{
    \institution{Imperial College London}
    \country{}
}

\author{Ariah Klages-Mundt}
\affiliation{
    \institution{Cornell University}
    \country{}
}
\author{Dominik Harz}
\affiliation{
    \institution{Interlay}
    \country{}
}

\author{William J. Knottenbelt}
\affiliation{
    \institution{Imperial College London}
    \country{}
}

%%For page numbers
\thispagestyle{plain}
\pagestyle{plain}

\begin{abstract}
Decentralized Finance (DeFi), a blockchain powered peer-to-peer financial system, is mushrooming.
Two years ago the total value locked in DeFi systems was approximately $700$m USD, now, as of April 2022, it stands at around $150$bn USD.
The frenetic evolution of the ecosystem has created challenges in understanding the basic principles of these systems and their security risks.
In this Systematization of Knowledge (SoK) we delineate the DeFi ecosystem along the following axes: its primitives, its operational protocol types and its security.
We provide a distinction between technical security, which has a healthy literature, and economic security, which is largely unexplored, connecting the latter with new models and thereby synthesizing insights from computer science, economics and finance.
Finally, we outline the open research challenges in the ecosystem across these security types.
\end{abstract}


\maketitle

\input{content/1_introduction}

\input{content/2_building_blocks}

\input{content/3_defi_protocols}

\input{content/3.1_exploits}

\input{content/4_technical_security}

\input{content/5_economic_security}

\input{content/6_challenges}

\input{content/7_conclusion}

\balance

\begin{acks}
We thank the anonymous reviewers for their feedback and suggestions.
This project received partial funding from EPSRC Standard Research Studentship (DTP) (EP/R513052/1), the Ethereum Foundation, the Brevan Howard Centre for Financial Analysis, Smart Contract Research Forum, and a Bloomberg Fellowship. 
\end{acks}

\hidelinks
\bibliographystyle{splncs04}
%\bibliography{references}

\input{references.bbl}
\vspace{4em}
\input{content/appendix}

\end{document}

%% file: content/1_introduction.tex
\section{DeFi: Finance 2.0?}
\label{sec:introduction}

Consider two views on the promise of Decentralized Finance (DeFi). 
For the DeFi Optimist, DeFi amounts to a breakthrough technological advance, offering a new financial architecture that is non-custodial, permissionless, openly auditable, (pseudo)anonymous, and with potentially new capital efficiencies.
According to this view, DeFi generalizes the promise at the heart of the original Bitcoin whitepaper~\cite{nakamoto08bitcoin}, extending the innovation of non-custodial transactions to complex financial operations.
In contrast, the DeFi Pessimist is concerned that, inter alia, the unregulated, hack-prone DeFi ecosystem serves to facilitate unfettered and novel forms of financial crime.
The pseudo-anonymous nature of DeFi permits cryptocurrency attackers, scammers, and money launderers to move, clean, and earn interest on capital.
A critical part of the debate between the DeFi Optimist and the DeFi Pessimist, but outside of the scope of this paper, is \emph{moral} in nature.
This SoK leaves this important facet aside, focusing instead on synthesizing and evaluating the technical innovations of DeFi, seeking to allow newcomers to the field to discover the essential features and problems of the DeFi terrain.

DeFi, in its ideal form, exhibits four properties. DeFi is:
\begin{enumerate}
    \item Non-custodial: participants have full control over their funds at any point in time
    \item Permissionless: anyone can interact with financial services without being censored or blocked by a third party
    \item Openly auditable: anyone can audit the state of the system, e.g., to verify that it is healthy
    \item Composable: its financial services can be arbitrarily \textit{composed} such that new financial products and services can be created (similar to how one is able to conceive new Lego models based on a few basic building blocks)
\end{enumerate}

% Traditional finance is based on a custodial model: banks hold custody of funds, stocks are held at a custodian bank, and collateral of contracts may be held in escrow accounts by a custodian. 
% For better or worse, these custodians have to be trusted and they need to be compensated for their custodial services. 
% In contrast, blockchain mechanisms provide a means for agents who do not trust each other to cooperate without requiring trusted third parties. 
% Holding on-chain assets can be done without a custodian, and general scripting functionality (``smart contracts'') can execute deterministically and verifiably on an underlying blockchain. 
% Among many uses, this allows collateral to be escrowed on-chain without a custodian, which opens up a variety of non-custodial applications.
% This allows capital to be seamlessly rehypothecated while following the protocol collateralization rules.

DeFi has grown rapidly, going from around $700$m USD in total value locked (TVL, or analogously ``assets under management'') at the start of $2020$ to over $150$bn USD as of April 2022.
Ethereum alone accounts for $75$bn USD in TVL, with the most capitalized use cases being collateralized lending, constituting c.$54$\%, of the TVL, and decentralized exchange (DEXs), constituting c.$31$\% of the TVL as of April 2022~\cite{web:defipulse}.
In turn, this rise led to the $24$ hour volume on a decentralized cryptoasset exchange~\cite{web:uniswap}, overtaking that of a major centralized cryptoasset exchange~\cite{web:coinbase} for the first time~\cite{web:uniswaptopplescoinbase}.

Yet, as with any nascent technology, DeFi is not without its risks.
% In the last year alone, DeFi has experienced more than $20$ major protocol exploits, resulting in a loss of funds amounting to over $130$m USD~\cite{defihacks}.
The decentralized nature of DeFi necessarily allows any actor to write \textit{unaudited} and even malicious smart contracts, where user funds can be lost through programming error or stolen.
Moreover, the audit process itself is no guarantee of safety, with many \textit{audited} protocols (e.g.,~\cite{hack:bzx,hack:akropolis,hack:cover,hack:harvest,hack:origin}) suffering serious exploits.
% Even at the technical level, the blockchains underlying DeFi are facing significant challenges, such as rising transaction costs that price out small transactions, in turn restricting the set of transaction types for which the layer-one blockchain can be used.
% Blockchain transaction fees have risen considerably during periods of congestion, with the fees for relatively simple smart contract operations running into the hundreds of dollars.
% Rising transaction costs price out small transactions, in turn restricting the set of transaction types for which the layer-one blockchain can be used.

\paragraph{This paper.}
For DeFi to fulfill the vision of the DeFi Optimist, it must first be secure.
The central contribution of this SoK is to cleanly and exhaustively delineate the DeFi security challenge into technical security and economic security.
The delineation centers on atomicity: whether the attack is near-instantaneous and can costlessly fail (and therefore risk-free), or has a non-instantaneous duration and where failure comes with a cost.
This categorization has the benefit of cleanly mapping different types of models to each type of security and clarifying previously vague terms of ``economic risk'' that have been commonly misapplied to exploits that are better understood as technical in nature (e.g., DEX sandwich attacks).
Prior to this paper, economic security risks were largely unexplored, in part because they require synthesizing insights from across computer science, economics, and finance.

This SoK is structured as follows.
We outline DeFi primitives in Sec.~\ref{sec:blockchain} and systematize existing DeFi protocols by six types of operations in Sec.~\ref{sec:defi-protocols}.
We provide a definition of DeFi exploits in Section~\ref{sec:exploits}.
We then define a novel functional categorization of technical and economic security risks in DeFi and classify different attacks for each security type in Sections~\ref{sec:technical-security} and~\ref{sec:economic-security}, respectively.
We then propose a set of six primary open research challenges for DeFi going forward that build on these security types in Sec.~\ref{sec:challenges}.

\paragraph{Related work.}
Several surveys and other SoKs exist on specific DeFi protocol types.\footnote{Most have been pre-printed since our paper was originally released.}
We direct the reader, when appropriate, to such SoKs for further material (e.g., \cite{bartoletti2020sok,cousaert2021sok}).
One well-structured survey on DeFi has been published after our pre-print \cite{schar2021decentralized},
which categorizes DeFi protocols but focuses on high level risk and regulatory challenges.
None of these works delineate the new types of security challenges across DeFi, which is the main focus of our SoK.

% \paragraph*{This Work}
% We outline the primitives for DeFi in Sec.~\ref{sec:blockchain} and then make the following contributions:
% \begin{itemize}
%     \item \textbf{Protocol Systematization:} We systematize the existing DeFi protocols according to six types of operations (Sec.~\ref{sec:defi-protocols}).
%     \item \textbf{Technical Security:} We define technical security in the context of DeFi as a risk-free earning potential and classify the set of technical attacks into three distinct categories. Technical security challenges such as smart-contract vulnerabilities serve to undermine the soundness of the ecosystem, limiting the extent to which it can be entrusted with funds (Sec.~\ref{sec:technical-security}).
%     \item \textbf{Economic Security:} We define economic security in the context of DeFi as secure incentive alignment of agents and organize the set of economic attack vectors into four distinct categories. The economic security risks emerge as the incentive mechanisms encoded in the underlying smart contracts make contact with reality (Sec.~\ref{sec:economic-security}).
%     \item \textbf{Holistic Security:} The distinction between technical and economic security illustrates that the development of the DeFi ecosystem is on two fronts. We synthesize insights from both these perspectives and propose a set of six primary open research challenges for DeFi going forward (Sec.~\ref{sec:challenges}). 
% \end{itemize}

%% file: content/2_building_blocks.tex
\section{DeFi Primitives}
\label{sec:blockchain}
DeFi protocols require an underlying distributed ledger such as a blockchain, a peer-to-peer distributed append-only record of transactions.
We take the underlying distributed ledger layer solely as an input into DeFi and refer the reader to existing work (notably~\cite{bonneau2015sok,bano2019sok,Gudgeon2019SoK,zamyatin2019sok}) for a fuller exposition of the blockchain layer itself.
We assume that the ledger has the basic security properties of consistency, integrity and availability~\cite{zhang2019security}.
Without these security properties, DeFi protocols built on top of such a ledger would themselves become inherently insecure.

In this section, we draw attention to features of the underlying blockchain layer that are most pertinent to DeFi.

\subsection{Smart Contracts and Transactions}
\label{sec:smart-contracts}

\paragraph{Smart Contracts}

The most important provision is that the underlying ledger offers the ability to use smart contracts, which are program objects that live on the blockchain.
These are able to communicate with one-another, via message-calls, within the same execution context and support \emph{atomicity}, i.e., a transaction either succeeds fully (state update) or fails entirely (state remains unaltered), such that no execution can result in an invalid state.
In some cases, \emph{bundles} of transactions, in which transactions are grouped together in a given sequence, can also be executed atomically (e.g., \cite{flashbotsBundles}).

% These are programs that encode a set of rules for processing transactions which are enforced by a blockchain's consensus rules, allowing for trustless economic interactions between parties. 
Smart contracts rely on blockchains that are transaction-based state machines, whereby an agent can interact with smart contracts via transactions.
Once a transaction is confirmed, the contract code is run by all nodes in the network and the state is updated.
The underlying cost to state updates comes in the form of transaction fees charged to the sender.
For instance, the Ethereum Virtual Machine (EVM)~\cite{wood2014ethereum} on the Ethereum~\cite{buterin2014next} blockchain is a stack machine which uses a specific set of instructions for task execution.
The EVM maintains a fixed mapping of how much gas, an Ethereum-specific unit that denominates computational cost, is consumed per instruction.
The total amount of gas consumed by a transaction is then paid for by the sender~\cite{perez2019broken,werner2020step}. 

In order for DeFi protocols to function on top of them, smart contracts must:
\begin{itemize}
    \item be expressive enough to encode protocol rules
    \item allow conditional execution and bounded iteration
    \item be able to communicate with one-another, via message-calls, within the same execution context (typically a transaction)
    \item support atomicity, i.e., a transaction either succeeds fully (state update) or fails entirely (state remains unaltered), such that no execution can result in an invalid state
\end{itemize}
These properties provide \textit{composability}, where smart contracts can be snapped together like Lego bricks (``Money Lego"~\cite{web:defi-pulse}), with the possibility of building complex financial architectures. % FIXME: missing citation~\cite{moneylegospapers}.
This is similar to as was envisaged in~\cite{jones2000composing}.
While promising, the side-effects of smart contracts interactions and the space of all possible interactions is vast.
In a setting focused on financial applications, such complexity brings with it a great burden to understand the emergent security properties of composed smart contracts.
We discuss this in more detail in Sections~\ref{sec:technical-security} and~\ref{sec:economic-security}.
One particularly common use of smart contracts is to implement tokens, on-chain assets.

% \paragraph{Tokens}
% \label{sec:tokens}
% A common use of smart contracts is to implement \mbox{\textit{tokens}}, which can be used to represent assets, ranging from Ether~\cite{web:weth} and other cryptoassets~\cite{web:wbtc} to synthetic assets or derivatives~\cite{web:synthetix}, as well as provide some utility, such as the right to participate in an election.
% Tokens are implemented by contracts adhering to a standard token interface, allowing protocols to easily handle different tokens without having to know about their implementation in advance.
% In Ethereum, tokens are usually implemented via the standardized ERC-20~\cite{web:eip20} and ERC-721~\cite{web:eip721} interfaces for fungible and non-fungible tokens, respectively~\cite{frowis2019detecting}, although other token standards exist~\cite{web:eip777,web:eip1155,web:eip1363}.
% A common distinction is between fungible tokens, which are interchangeable~\cite{web:eip20}, and non-fungible tokens which are distinct~\cite{web:eip721}.

\paragraph{Transaction Execution}
\label{sec:transaction-execution}
When a blockchain network participant wishes to make a transaction, the details of the unconfirmed transaction are first broadcast to a network of peers, validated, and then stored in a waiting area (the \textit{mempool} of a node).
This mempool is then propagated among the network nodes.
Participants of the underlying ledger responsible for ensuring consensus, \textit{miners}, then choose which transactions to include in a given block, based in part on the transaction fee attached to each transaction.
% Transactions in a block are executed sequentially in the order in which the \textit{miner} of the respective block included them. 
% For a detailed treatment of how this process works, we refer the reader to~\cite{nakamoto08bitcoin,wood2014ethereum,narayanan2016bitcoin,10.1145/3419394.3423628}.
% Miners ability to control the sequence in which transactions are executed means that miners can order transactions in ways that will earn them revenues.
% In addition, they can insert their own transactions to extract further revenues.
% Miners can even be bribed to undertake such transaction re-ordering~\cite{mccorry2018smart,winzer2019temporary}.
% The value that miners can extract is known as \textit{Miner Extractable Value} (MEV)~\cite{daian2019flash}.
% We consider these issues in detail in Sec.~\ref{sec:mev}.
Transactions in a block are executed sequentially in the order in which the \textit{miner} of the respective block included them. 
For a detailed treatment of how this process works, we refer the reader to~\cite{nakamoto08bitcoin,wood2014ethereum,narayanan2016bitcoin,10.1145/3419394.3423628}.
Miners have the ability to control the sequence in which transactions are executed.
Hence, miners can order transactions in ways that will earn them revenues and even insert their own transactions to extract further revenues.
Miners can even be bribed to undertake such transaction re-ordering~\cite{mccorry2018smart,winzer2019temporary}.
The value that miners can extract is known as \textit{Miner Extractable Value} (MEV)~\cite{daian2019flash}.
We consider these issues in detail in Sec.~\ref{sec:mev}.

\subsection{Keepers}
\label{sec:keepers}
Protocols may rely on their on-chain state being continually updated for their security.
In transaction-based systems, updating the on-chain state requires transactions that are triggered externally.
Since smart contracts are not able to create transactions programmatically, protocols must rely on external entities to trigger state updates.
These entities, \textit{keepers}, are generally financially incentivized to trigger such state updates.
For instance, if for whatever reason a protocol requires a user's collateral to be automatically liquidated under certain conditions, the protocol will incentivize keepers to initiate transactions to trigger such liquidation.

\subsection{Oracles}
\label{sec:oracles}
An oracle is a mechanism for importing off-chain data into the blockchain virtual machine so that it is readable by smart contracts. 
This includes off-chain asset prices, such as ETH/USD, as well as off-chain information needed to verify outcomes of prediction markets.
Oracles are relied upon by various DeFi protocols (e.g.~\cite{leshner2019compound,whitepaper:aave,whitepaper:maker,synthetix2020litepaper,peterson2015augur}).

Oracle mechanisms differ by design and their risks, as discussed in~\cite{klages2020stablecoins,liu2020look}.
A centralized oracle requires trust in the data provider and bears the risk that the provider behaves dishonestly should the reward from supplying manipulated data be more profitable than from behaving honestly.
Decentralized oracles offer an alternative.
As the correctness of off-chain data is not verifiable on-chain, decentralized oracles tend to rely on incentives for accurate and honest reporting of off-chain data.
However, they come with their own shortcomings.
We provide a detailed overview of oracle manipulation risks and on the shortcomings of on and off-chain oracles in Sections~\ref{sec:single-tx-attacks} and~\ref{sec:market-manipulation}.

\subsection{Governance}
\label{sec:governance}
Governance refers to the process through which a system is able to effect change to the parameters which establish the terms on which interactions between participants within the system take place~\cite{klages2020stablecoins}.
Such changes can be performed either algorithmically or by agents.
While there is existing work on governance in relation to blockchains more broadly (e.g.~\cite{reijers2016governance,beck2018governance,lee2020political}), there is still a limited understanding of the properties of different mechanisms that can be used both for blockchains and DeFi.

Presently, a common design pattern for governance schemes is for a DeFi protocol to be instantiated with a benevolent dictator ---sometimes distributing power over a small council or ``multisig''--- who has control over governance parameters, with a promise made by the protocol to eventually decentralize its governance process.
Such decentralization of the governance process is most commonly pursued through the issuance of a governance token (e.g.~\cite{web:compoundfinance,web:maker,web:curve,web:balancer}), an ERC-20 token which entitles token holders to participate in protocol governance via voting on and possibly proposing protocol updates. This token represents ownership in a decentralized autonomous organization (DAO) that is taksed with stewardship of the protocol.

Protocol upgrades in the DAO setting come through proposals in the form of executable code, on which governance token holders vote.
In order to propose protocol updates, the proposer has to hold or have been delegated a threshold number of governance tokens.
For a protocol upgrade to be executed, a minimum number of votes is required, which is commonly called a ``quorum'' in this setting.
%We return to governance in Sec.~\ref{sec:economic-security}.

%% file: content/3_defi_protocols.tex
\section{DeFi Protocols}
\label{sec:defi}

\label{sec:defi-protocols}

\begin{figure}[htp]
    \centering
    \includegraphics[width=1\columnwidth]{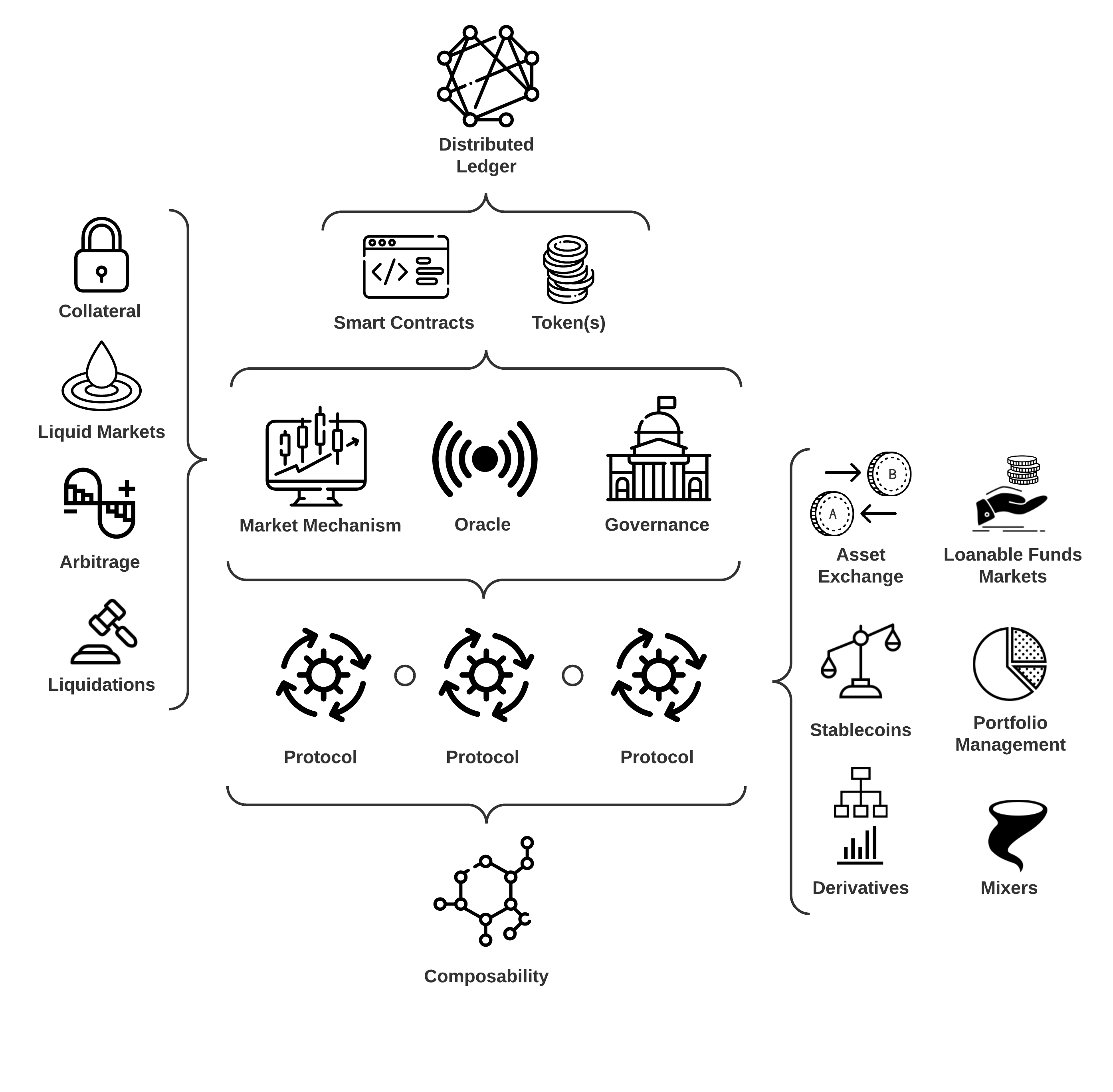}
    \caption{A conceptual overview of the different constructs within the DeFi ecosystem.}
    \label{fig:defi-overview}
\end{figure}

We now present DeFi protocols categorized by the type of operation they provide (for an illustration see Fig.~\ref{fig:defi-overview}).

% An overview is shown in Figure~\ref{fig:defi-overview}, while a classification of a selection of existing DeFi protocols is given in Appendix~\ref{appendix:defi-protocols}.

% \begin{figure}[htp]
%     \centering
%     \includegraphics[width=0.75\columnwidth]{figures/overview-defi-system-full-2.pdf}
%     \caption{A conceptual overview of the different constructs within the DeFi ecosystem.}
%     \label{fig:defi-overview}
% \end{figure}

\subsection{On-chain Asset Exchange}

Decentralized exchanges (DEXs)~\cite{lin2019deconstructing,web:index-dexs} are a class of DeFi protocol that facilitate the non-custodial exchange of digital assets, where all trades are settled on-chain and thus publicly verifiable. 
%thereby ensuring public verifiability for all transactions to network participants.
While DEXs initially only supported assets native to the chain on which they operate, wrapped tokens, such as wBTC~\cite{web:wbtc} (wrapped Bitcoin), and novel cross chain solutions~\cite{zamyatin2019xclaim,zamyatin2019sok,dubovitskaya2021gametheoretic,xu2020gametheoretic} have enabled DEXs to overcome this limitation. 
Today, based on the mechanism for price discovery, DEXs come in different variants, such as \textit{order book DEXs} (including individual~\cite{idex-whitepaper,warren20170x} and batch settlement~\cite{benes2017dutchx,gnosis}, see Appendix~\ref{app:batch-settlement-gnosis} for the latter) and \textit{automated market makers} (AMMs) (e.g.,~\cite{egorov2019stableswap,whitepaper:uniswap,whitepaper:balancer}).
Due to their widespread adoption and novelty in DeFi, we specifically focus on AMMs.

In traditional finance, market makers are liquidity providers that both quote a bid and ask price, selling from their own book, while making a profit from the bid-ask spread.
Optimal market making strategies quickly become sophisticated optimization problems.
In contrast, AMMs provide liquidity algorithmically through simple pricing rules with on-chain liquidity pools in place of order books and have been previously studied in algorithmic game theory, e.g., logarithmic market scoring rule (LMSR) \cite{hanson2003combinatorial} in prediction markets.
While they have largely remained unimplemented in traditional finance, they have become popular in DeFi for a several reasons:
(1) they allow easy provision of liquidity on minor assets,
(2) they allow anyone to become a market maker, even if the market making returns are suboptimal,
(3) AMM pools can be separately useful as automatically rebalancing portfolios,
(4) maintaining an order book on-chain is inefficient.

In an AMM liquidity pool, reserves for two or more assets are locked into a smart contract, where for a given pool, each liquidity provider receives newly minted liquidity tokens to represent the share of liquidity they have provided.
A trade is then performed by trading against a smart contract's liquidity reserve for an asset, whereby liquidity is added to the reserves of one token and withdrawn from the reserves of one or more other pool tokens.
A trading fee is retained by a liquidity pool and paid out proportionally to the amount of liquidity provided by each liquidity token holder.
Liquidity providers are required to give up their liquidity tokens in order to redeem their share of liquidity and accrued fees.

With an AMM, the price of an asset is deterministic and decided by a formula, not an order book, and thus depends on the relative sizes of the provided liquidity on each side of a currency pair. 
If the liquidity is thin, a single trade can cause a significant fluctuation in asset prices relative to the overall market, and arbitrageurs can profit by closing the spread.
Arbitrage refers to the process of buying or selling the same asset in different markets to profit from differences in price. 
Parties who undertake this process are \emph{arbitrageurs}, and play a critical role in DeFi protocols.
Arbitrage is used to ensure that the price for an asset on an AMM is at parity with the price on the open market.
Note that as the reserve ratios for a pool's assets change as liquidity is added and withdrawn, a liquidity provider may receive a different token ratio upon withdrawing his liquidity share compared to the ratio he initially deposited.
For a more focused and formal analysis of AMM design and the underlying market making mechanism, we direct the reader to~\cite{angeris2020improved,angeris2019analysis,angeris2021replicating,zhang2018formal,angeris2020does}.

\subsection{Loanable Funds Markets for On-chain Assets}
\label{sec:plfs}

Lending and borrowing of on-chain assets is facilitated through \textit{protocols for loanable funds} (PLFs)~\cite{gudgeon2020defi}, which refer to DeFi lending protocols that establish distributed ledger-based markets for loanable funds of cryptoassets by pooling deposited funds in a smart contract.
In the context of a PLF, a \textit{market} refers to the total supplied and total borrowed amounts of a token, where the available deposits make up a market's liquidity.
An agent may directly borrow against the smart contract reserves, assuming the market for the token is sufficiently liquid, where the cost of borrowing is given by the market's interest rate.
% The cost of borrowing is given by an interest rate, which is determined by a market's underlying interest rate model~\cite{xu2021banks}. 

On PLFs, loans are generally of two forms: \textit{over-collateralized loans} and \textit{flash loans}.
With an over-collateralized loan, a borrower is required to post collateral, i.e., provide something of value as security to cover the value of the debt, where the value of the collateral posted exceeds the value of the debt.
In this way, collateralization simultaneously ensures that the lender (likely a smart contract) can recover their loaned value and provides the borrower with an incentive to repay the loan.
% The idea is to ensure that even if the value of the collateral relative to the debt falls considerably, there would still be sufficient collateral to cover the debt.
In case the value of the locked collateral falls below some liquidation threshold, so-called liquidators, a type of keeper, are able to purchase the locked collateral at a discount and close the borrower’s debt position~\cite{perez2020liquidations}.
%In a liquidation scenario, the liquidated borrower would receive the collateral minus any outstanding debt and incurred penalty charges~\cite{perez2020liquidations}.

An alternative to over-collateralized loans are flash loans.
These are uncollateralized loans for the duration of a single transaction, requiring the borrower to repay the full borrowed amount plus interest by the end of the transaction.
Flash loans leverage a blockchain's atomicity (i.e., the transaction fails if the loan is not repaid in the same transaction) and offer several use cases, such as decentralized exchange arbitrage and collateral swaps.
However, they can also be used in attacks~\cite{qin2020attacking}.
For a more detailed discussion and formal analysis of PLFs, we direct the reader to \cite{gudgeon2020defi,bartoletti2020sok}.

\subsection{Stablecoins}

Non-custodial stablecoins are cryptoassets which aim to be price stable relative to a target currency, commonly the USD, and seek to achieve this via additional economic mechanisms.
% As of the time of writing, there are about a dozen non-custodial stablecoins, of which perhaps the most notable is MakerDAO's Dai~\cite{whitepaper:maker}, which has close to $7.80$bn USD in market capitalization as of April $2021$\footnote{Source: \url{https://defipulse.com/}. Accessed: 07-04-2021.}.
Note that custodial stablecoins, such as USDT~\cite{whitepaper:tether} are not within the scope of DeFi, since these principally rely on a trusted third-party to operate, though they may be among the assets used in other DeFi protocols.

In the decentralized setting, the challenge for the protocol designer is to construct a stablecoin which achieves price stability in an economically secure and stable way and wherein all required parties can profitably continue to participate~\cite{klages2020stablecoins}.
Price-stability is pursued via the use of on-chain collateral, providing a foundation of secured loans from which the stablecoin derives its economic value. 

The core components of a non-custodial stablecoin are as follows~\cite{klages2020stablecoins}. 
\begin{itemize}
    \item Collateral. This is the store of primary value for a stablecoin. Collateral can be exogenous (e.g., ETH in Maker~\cite{web:maker}), where the collateral is primarily used externally to the stablecoin, endogenous (e.g., SNX in Synthetix~\cite{web:synthetix}), where the collateral was created to be collateral or implicit (e.g., Nubits~\cite{Nubits2014}), where the design lacks an explicit store of collateral. 
    \item Agents. Agents form at least two roles in a non-custodial stablecoin: (1) risk absorption, for instance by providing collateral that is intended to absorb price risk, and (2) stablecoin users.
    \item Governance. A mechanism and set of parameters that governs the protocol as a whole (either performed by agents or algorithmically).
    \item Issuance. A mechanism to control the issuance of stablecoins against or using the collateral (either performed by agents or algorithmically).
    \item Oracles. A mechanism to import data external to the blockchain onto the blockchain, such as price-feeds. 
\end{itemize}
See \cite{klages2020stablecoins,zhao2021understand} for a more complete discussion of stablecoin designs, models, and challenges.

\subsection{Portfolio Management}

For liquidity providers seeking to maximize their returns, liquidity allocation can be an onerous task given the complex and expansive space of yield-generating options.
The management of on-chain assets can thus be automated through DeFi protocols which serve as decentralized investment funds, where tokens are deposited into a smart contract and an investment strategy that entails transacting with other DeFi protocols (e.g., PLFs) is encoded in the contract.
Yield in DeFi is generated through interest (including accrued fees earned) and token rewards. For the latter, a protocol (e.g., PLF or AMM) distributes native tokens to its liquidity providers and/or users as rewards for the provision of deposits and/or protocol adoption.
These protocol-native token rewards are similar to equity in the sense that they serve as a right to participate in the protocol's governance, as well as often represent a claim on protocol-generated earnings.
The distribution model for token rewards in exchange for supplied liquidity may vary across protocols, yet is commonly proportional to how much liquidity an agent has supplied on a protocol.
Therefore, smart contract-encoded investment strategies of on-chain assets are tailored around yield generating mechanisms of different protocols with the sole aim of yield aggregation and maximization.
In practice, on-chain management of assets may range from automatic rebalancing of a token portfolio~\cite{feng2019set} to complex yield aggregating strategies~\cite{web:yearnfinance}.

\subsection{Derivatives}
\label{sec:derivatives}

Derivatives are financial contracts which derive their value from the performance of underlying assets.
As of March 2022, the derivatives market represents about $62$\% of the entire cryptoassets trading market~\cite{web:cryptocompare}.
While about $99$\% of the derivative trading volume is achieved on centralized exchanges, a number of DeFi protocols have emerged which provide similar functionality~\cite{web:opyn,web:dydx,web:opyn,wintermute2020hegic}, with a particular focus on synthetic assets, futures, perpetual swaps and options.
We lay out the adoption four different basic types of derivatives popular in the cryptoasset space\footnote{For an introduction to derivatives, we refer the reader to~\cite{hull2009options}}:

\begin{enumerate}
\item \point{Synthetic assets}
In DeFi, synthetic assets typically replicate off-chain assets on-chain (e.g., the USD in protocols like Maker and Synthetix~\cite{web:synthetix}).
Though less used at present, another mechanism for constructing synthetic assets is to use AMMs that enact dynamic portfolio rebalancing strategies to replicate derivative payoffs. 
These bear a resemblance to synthetic portfolio insurance (see Ch.~$13$ in~\cite{hull2009options}) in traditional finance and have been explored more specifically using constant product market makers in~\cite{clark2020replicating,evans2020liquidity}.
\item \point{Futures}
Futures have seen little adoption in DeFi yet.
Likely this is caused by the high volatility of the underlying cryptoassets making it hard to determine the risk taken by traders writing the futures.
\item \point{Perpetual Swaps}
These are similar to futures, however, they have no set expiry date or settlement and were specifically created and popularized for cryptoasset markets~\cite{web:bitmex}.
Perpetuals allow traders to decide (typically on a daily basis, e.g.,~\cite{web:dydx}) to keep the position by providing a funding transaction in case their position is underfunded.
Due to the frequent price discovery, the price of perpetuals trades typically closer to the underlying in comparison to futures.
Moreover, perpetuals are more capital efficient than trading the underlying itself since platforms require less than 100\% collateral be posted by traders.
\item \point{Options}
Currently, the DeFi market for options is very early with basic call and put options (e.g.,~\cite{web:opyn,wintermute2020hegic}).
The cause for the limited adoption of options is three-fold.
First, current option platforms are at least $100$\% collateralized.
In comparison to their centralized counter-parts, this represents large capital inefficiency.
Second, derivatives with set expiry dates like futures and options are hard to price on AMMs.
Most AMM platforms (e.g., Uniswap~\cite{web:uniswap}) do not account for a time dimension in the asset.
This causes an issue specifically with option trading since the value of the option is subject to time decay.
Possible remedies are more nuanced AMM designs like \cite{yieldspace} that aim to incorporate such a time dimension.
Also, complex value functions in the AMM like Balancer~\cite{web:balancer} allow replicating strategies that combine the underlying and a derivative into a single asset~\cite{evans2020liquidity}.
Third, options require a liquid market for efficient price discovery.
Adoption will require solving the above problems to bootstrap the required liquidity that allows efficient pricing of those options.
\end{enumerate}

\subsection{Privacy-preserving Mixers}
\label{sec:mixers}
Mixers are methods to prevent the tracing of cryptocurrency transactions. These are important to preserve user privacy, as the transaction ledger is otherwise public information; however, this also means they could be used to obscure the source of illicit funds.
Mixers work by developing a ``shielded pool'' of assets that are difficult to trace back before entering the pool.
They typically take one of two forms: (1) mixing funds from a number of sources so that individual coins can't easily be traced back to address individually (also called a ``coinjoin'', e.g., \cite{wasabi}), or (2) directly shielding the contents of transactions using zero knowledge proofs of transaction validity (e.g., \cite{tornado,zcash}).
Mixers serve as a DeFi-like application itself and additionally as a piece that could be included within other DeFi protocols.

%% file: content/3.1_exploits.tex
\section{What is a DeFi exploit?}
\label{sec:exploits}

We define what we consider an exploit to be in the DeFi context.
To do this, we first need to set out a taxonomy of blockchain information.
Let $t$ denote the ordering of events. 
At any point $t$ in this sequence of events, there are three categories of information relevant to a DeFi protocol:
\begin{itemize}
    \item Off-chain ground truth, denoted $\mathcal{G}_t^{OFF}$. For example, true wind speed or the true equilibrium prices of assets traded in off-chain venues;
    \item On-chain ground truth, denoted $\mathcal{G}_t^{ON}$. For example, on-chain asset ownership;
    \item On-chain estimates of off-chain ground truth, denoted $\hat{\mathcal{G}}_t^{OFF}$. For example, oracle reported prices.
\end{itemize}

A smart contract only has access to $\mathcal{G}_t^{ON}$ and $\hat{\mathcal{G}}_t^{OFF}$.
Now consider that a set of smart contracts (e.g., constituting a DeFi protocol), $S$, has a set of intended properties, $P^S$.
Each intended property is a function of the information up until the current state in the sequence such that
$$P^S \left(\mathcal{G}^{ON}_t, \mathcal{G}^{OFF}_t \right) = \left\{ P_0^S \left(\mathcal{G}^{ON}_t, \mathcal{G}^{OFF}_t \right), \ldots, P_n^S \left(\mathcal{G}^{ON}_t, \mathcal{G}^{OFF}_t \right) \right\}.$$
For a particular property, it can either hold or not hold at any given state, as intended, such that the properties are Boolean.
Some examples of these properties include the achievement of an invariant rule for an AMM or ensuring that debt positions remaining over-collateralized in accordance with a collateral factor in a Protocol for Loanable Funds (PLF).

For a smart contract to execute as intended, first, the smart-contract program must actually implement the intended logic (i.e., it must be free of implementation bugs).
Second, since smart contracts often heavily depend on $\mathcal{G}_t^{OFF}$, on-chain estimates of off-chain information, $\hat{\mathcal{G}}_t^{OFF}$, must be sufficiently accurate to ensure that $P_i (\mathcal{G}_t^{ON}, \mathcal{G}_t^{OFF}) = P_i (\mathcal{G}_t^{ON}, \hat{\mathcal{G}}_t^{OFF})$.
Such estimates can be manipulated: for example, smart contract states can experience `flash' manipulation, where an artificial state is set up at the beginning of a transaction and unwound at the end of a transaction.
Such manipulation can result in $\mathcal{G}_t^{OFF}$ and $\hat{\mathcal{G}}_t^{OFF}$ diverging.

The different categories of exploit now follow from this statement of what is required for a smart contract to reliably execute as intended. An exploit can arise from an attacker gaining, or seeking to gain from:
\begin{enumerate}
    \item Seizing upon a difference between the actual implementation of a smart contract and the intended implementation. See Section~\ref{sec:smart-contract-vulnerabilities}.
    \item Seizing upon an incidental difference between $\mathcal{G}_t^{OFF}$ and $\hat{\mathcal{G}}_t^{OFF}$. For example, in November 2020 an incidental deviation in the Dai price reported by Coinbase triggered liquidations on Compound\cite{web:cointelegraph-compound}.\footnote{Where such incidental differences are not seized upon this might more properly be considered a system failure, but we group these together as both are relevant for analyzing protocol security}
    \item Creating a deliberate difference between $\mathcal{G}_t^{OFF}$ and $\hat{\mathcal{G}}_t^{OFF}$. For examples, see \ref{sec:single-tx-attacks} and \ref{sec:tx-ordering-attacks}.
\end{enumerate}
In exploits of types (2) and (3), a protocol's operation is made to deviate from its intended properties, i.e., for some $i, t$, 
$$P_i \left(\mathcal{G}_t^{ON}, \mathcal{G}_t^{OFF}\right) \neq P_i \left(\mathcal{G}_t^{ON}, \hat{\mathcal{G}}_t^{OFF}\right).$$

Note the importance of the notion of intention in this delineation of exploit types. 
For example, when prices on an AMM and prices off-chain differ, this does not amount to an exploit because an intended property of an AMM is that arbitrage will rebalance the AMM pool according to the AMM's pricing rule.
However, when a protocol uses an AMM price as a price oracle, its operation is relying on information from $\hat{\mathcal{G}}_t^{OFF}$. When this information is significantly different from the true information in $\mathcal{G}_t^{OFF}$ (e.g., if an attacker manipulates the oracle price), this can cause the protocol to behave in unintended ways, which would amount to an exploit.

Exploits are usually worrisome when they are either profitable (i.e., an attacker can get more assets out of the exploit than they are spending to execute the exploit, usually from stealing assets) or when they cause high losses for protocols or users for low cost of execution (which itself is often profitable by short selling the token of a protocol to be exploited).

In the following two sections we delineate between two general types of exploits in a novel way that simultaneously distinguishes how the exploit is performed, the type of risks taken on by an attacker, and the types of tools and models required to analyze security in the two contexts.
Note that some attacks on DeFi protocols may not be categorized as security exploits at all. For example, the collapse of the Terra stablecoin \cite{bloomberg2022Terra} (and related collapses like the Iron stablecoin before it \cite{defiant2021Iron}) may be better described as currency runs/attacks as opposed to security exploits considering that they result from breaching the economic limits of a mechanism without necessarily exploiting the formal properties of smart contracts. 

%% file: content/4_technical_security.tex
\section{Technical Security}
\label{sec:technical-security}

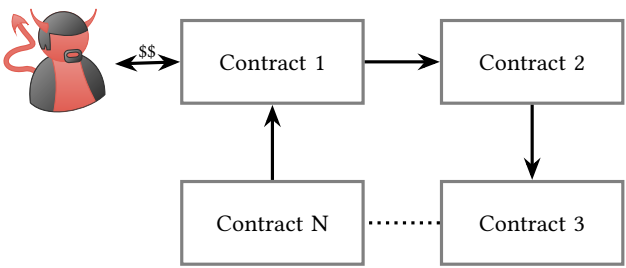
\begin{figure}[htp]
  \centering
      \begin{tikzpicture}[auto,
                     > = Stealth, 
         node distance = 10mm,
            box/.style = {draw=gray, very thick,
                          minimum height=11mm, text width=22mm, 
                          align=center},
     every edge/.style = {draw, very thick},
every edge quotes/.style = {font=\footnotesize, align=center, inner sep=1pt}
                          ]
% from bottom to top
  \node (n11) [devil,evil,minimum size=1cm] {};
  \node (n12) [box, right=of n11] { Contract 1};
  \node (n13)[box, right=of n12] {Contract 2};
  \node (n21) [box, below=of n12] {Contract N};
  \node (n22) [box, below=of n13] {Contract 3};
%Lines
\draw
      (n11) edge [<->, "\$\$"]        (n12)
      (n12) edge [->]        (n13)
      (n13) edge [->]        (n22)
      (n22) edge [dotted]    (n21)
      (n21) edge [->]        (n12);      
      \end{tikzpicture}
      \caption{Diagram of a technical exploit.}
      \label{fig:technical-security-illustration}
\end{figure}

We define a DeFi security risk to be \emph{technical} if an agent can atomically exploit a protocol.
In a technical exploit, an attacker effectively finds a sequence of contract calls, whether in a single transaction or a bundle of transactions, that leads to a profit through a violation of a protocol's intended properties (as visualized in Fig.~\ref{fig:technical-security-illustration}).
Such exploits can be performed risk-free (and often in a sense `instantaneously') because the outcomes for the attacker are binary: either the attack is successful and the attacker profits or the transaction reverts (effectively the attack doesn't happen) and the attacker only loses some gas fees.

In current blockchain implementations, this coincides with (1) manipulating an on-chain system within a single transaction, which is risk-free for anyone, and (2) manipulating transactions within the same block, which is risk-free for the miner generating that block or for an attacker who creates a bundle of transactions that are required to execute atomically in the order given.
By exploiting technical structure, the underlying blockchain system allows no opportunity for markets or other agents to react in the course of such exploits (when such reaction is possible, we enter the domain of \emph{economic} security problems in the next section).
When there is competition to perform these exploits, they will factor into the game theory of blockchain mining (e.g., \cite{biais2019blockchain}) as part of MEV extraction (as discussed in \cite{daian2019flash}); however, attempting these exploits will be risk-free (minus potential gas fees) for any attacker.
We identify three categories of attacks that fall within technical security risks of DeFi protocols: attacks exploiting smart contract vulnerabilities, attacks relying on the execution order of transactions in a block, as well as attacks which are executed within a single transaction.
These security risks are often addressable with program analysis and formal models to specify protocols, although these problems can quickly become complex to formulate and computationally hard.

\begin{tcolorbox}[boxsep=1pt,left=2pt,right=2pt,top=2pt,bottom=2pt, title=Technical Security]
\emph{A DeFi protocol is technically secure if it is not possible for an attacker to atomically exploit the protocol at the expense of value held by the protocol or its users. Due to atomicity, these attacks can generate risk-free profit. A common property of technical exploits is that they occur within a single transaction or a bundle of transactions in a block.}
\end{tcolorbox}

Examples of past DeFi protocol exploits that fall into the presented attack categories of technical security are given in Table~\ref{tab:technical-attacks} in Appendix~\ref{appendix:defi-protocols}.
% We discuss a subset of these exploits as practical examples in the context of the attack category the exploit falls under.

\subsection{Smart Contract Vulnerabilities}
\label{sec:smart-contract-vulnerabilities}
%Smart contracts being at the center of any DeFi protocol, any vulnerabilities in their implementation can cause them to be at loss.
Smart contract vulnerabilities have been extensively discussed in the literature~\cite{atzei2017survey,perez2019smart,tsankov2018securify} and we will therefore not give an extensive list of all the known vulnerabilities but rather focus on the one which have already been exploited in the DeFi context.

\point{Reentrancy}
A contract is potentially vulnerable to a reentrancy attack if it delegates control to an untrusted contract, by calling it with a large enough gas limit, while its state is partially modified~\cite{tubiblio111410}.
A trivial example is a contract with a \lstinline{withdraw} function that checks for the internal balance of a user, sends them money and updates the balance.
If the receiver is a contract, it can then repeatedly re-enter the victim's contract to drain the funds.
Although this attack is already very well-known, it has been successfully used several times against DeFi protocols\cite{web:dforce,hack:imbtc-uniswap}.
%We briefly present two of these attacks in more detail.

% \textit{dForce:} One of the most prominent examples of this exploit was against the dForce protocol~\cite{web:dforce}, which features a PLF, in April $2020$ to drain around $25$ million USD worth of funds~\cite{web:dforce-hack}.
% The attacker used imBTC~\cite{web:imbtc}, which is an ERC-777 token~\cite{web:eip777}, to perform the attack.
% A particularity of ERC-777 tokens, as opposed to ERC-20 tokens, is that they have a hook calling the receiver when the receiver receives funds.
% This means that any ERC-777 tokens will indirectly result in the receiver having control of the execution.
% In the dForce attack, the attacker used this reentrancy pattern to repeatedly increase their ability to borrow without enough collateral to back up their borrow position, effectively draining the protocol's funds.

% \textit{imBTC Uniswap Pool:} Despite the fact that Uniswap does not support ERC-777 tokens~\cite{whitepaper:uniswap}, an imBTC Uniswap~\cite{web:uniswap} pool worth roughly $300\,000$ USD was drained using the reentrancy attack.

% Both of these attacks show a common attack pattern in DeFi applications: identifying and exploiting attack vectors which exploit protocols' interconnectedness, where the composability risks therein are often under-examined. 
In practice, reentrancy vulnerabilities are generally simple to detect and fix by using static analysis tools~\cite{tsankov2018securify,luu2016making}.
There are two main ways to prevent this vulnerability: (1) using a reentrancy guard that prevents any call to a given function until the end of its execution or (2) finalizing all the state updates before passing execution control to an untrusted contract.

\point{Integer Manipulation}
Almost every DeFi application manipulates monetary amounts in some way or another.
This often involves not only adding and subtracting to balances but also converting into different units or to different currencies.
We present the two most common types of integer manipulation issues.

The first issue, which has been extensively studied in the literature~\cite{10.1145/3274694.3274737,DBLP:conf/ndss/KalraGDS18}, is integer over- and underflow.
% The EVM does not raise any exception in case of over- or underflow and without correct checks, such overflows could stay undetected until the value is used in some sort of action such as, for example, a transaction sending a token amount.
As the Ethereum Virtual Machine (EVM)~\cite{wood2014ethereum} does not raise any exception in case of over- or underflow, this will often result in failed transactions and cause the smart contract to misbehave~\cite{perez2019smart}.

The second issue is unit error during integer manipulation.
% Limitations in the expressivity of both the programming language and the virtual machine, as well as poor development practices have caused issues related to this type of arithmetic operations.
The main language used to develop DeFi applications on Ethereum at the time of writing is Solidity~\cite{docs:solidity}, which has a limited type system and no support for operator overloading.
In addition, the EVM only supports a single type, $32$ byte integers, and has no built-in support for fixed-point numbers.
To work around this limitation, each protocol decides on an arbitrary power of $10$ to use as its base unit, often $10^{18}$, and all the computations are performed in terms of this unit.
However, given the limitations of the type-system, most programs elect to use exclusively $32$ byte integers.
Arithmetic on two units accidentally on different scales would not be caught by the compiler.
% These shortcomings can result in substantial losses in practice, as was the case for the Yam Finance protocol~\cite{web:yam-bug,web:yam-bug-analysis}, resulting in $750\,000$ USD worth of tokens~\cite{medium:yam_finance} to be locked indefinitely.

\point{Logical Bugs} 
There are a large number of exploits that are rooted in simple programming errors in the smart contracts.
While logical bugs are by no means unique to smart contracts, but common to any type of software, the consequences for smart contracts, where immutability underpins the system, can be much more severe than for many other genres of software and result in unrecoverable financial losses.
% One such logical bug that resulted in notable financial losses to highlight the often trivial nature of the issue encountered:

% Although this is only a single instance of smart contract logical bugs, 
A large share of the bugs found in~\autoref{tab:technical-attacks} are also very simple mistakes that have been overlooked in both the development process and professional contract audits.
We discuss in Section~\ref{sec:challenges} potential mitigation techniques to these issues.

\subsection{Single Transaction Attacks}
\label{sec:single-tx-attacks}
We refer to attacks which can be successfully executed, independent of knowing about some other pending transaction, as single transaction attacks.
This category of attack is leveraging transaction atomicity and composability of smart contracts.

\point{Governance Attacks}
Protocols that implement decentralized governance mechanisms tend to rely upon governance tokens, which empower token holders to propose and vote on protocol upgrades.
The technical structure of these governance mechanisms can sometimes be exploited.
For instance, many governance designs allow updates to be instantaneously proposed and approved and the protocol code upgraded within a single transaction.
In this setting, an attacker may obtain an amount of governance tokens sufficient to propose and execute malicious contract code and steal a contract's funds, all while circumventing the usual governance process in which other participants can react~~\cite{gudgeon2020decentralized}.
The attacker may even obtain the share of governance tokens instantaneously within the same transaction (i.e., they may not be a long-term participant) through flash loans from PLFs and swaps from AMMs.
In fact, large quantities of governance tokens can be obtained easily in these ways today, and such attacks have been executed in practice~\cite{makerdao2020flash}.

The direct problem can usually be sorted by applying a timelock to the governance process so that updates cannot be performed instantaneously and other participants have a chance to react.
However, as we will see in the economic security section, this often does not solve the incentive issues completely, it just resolves the \emph{technical} issue.

\point{Single Transaction Sandwich Attacks}
\label{sec:technical-sec:single-tx-sandwich}
In a single transaction sandwich attack, an attacker manipulates an instantaneous AMM price in order to exploit a smart contract that uses that price. 
An attacker first creates an imbalance in an AMM, exploits composable contracts which rely on the manipulated price, and then reverses the imbalance to cancel out the cost of the first step.
The whole sequence can be performed atomically in a single transaction risk-free.
Creating an imbalance typically requires access to a large amount of capital. 
In a system with flash loans/minting, all agents effectively have such access, although we stress that these attacks are still possible for large capital holders regardless of whether flash loans/minting are widespread.
In practice, this type of attack has occurred multiple times~\cite{peckshield2020value,rekt2020harvest}. 
One of the most prominent single transaction sandwich attacks in terms of seized funds was performed against the Harvest protocol~\cite{exploit:harvest-tx-analysis}.
The attacker took out a \$$50$m USDT flash loan from Uniswap and used part of the funds to create an imbalance in the liquidity reserves of USDC and USDT on Curve~\cite{web:curve} (an AMM) to increase the AMM's virtual price of USDT.
As the price of USDT on Curve was used as an on-chain oracle by the Harvest protocol, the attacker was able to mint Harvest LP tokens (i.e., tokens a liquidity provider receives in exchange for depositing funds into a protocol) by depositing $60.6$m USDT, before reversing the imbalance on Curve and withdrawing $61.1$m USDT from Harvest.
The attacker was able to withdraw more USDT than deposited, as at the time of the withdrawal, the USDT price given by Curve was less than the deposit price, and therefore one Harvest LP token was worth more USDT during withdrawal.
The attacker repeated this attack $32$ times, draining a total of \$$33.8$m of the protocol's funds.

To protect against such manipulations, AMMs include a limit amount (or maximum slippage) that a trade can incur, though this only prevents manipulations above this amount.
% The severity single transaction sandwich attacks occurring in practice is highlighted by the following example:

\subsection{Transaction Ordering Attacks}
\label{sec:tx-ordering-attacks}
In traditional finance, the act of \textit{front-running} refers to taking profitable actions based on non-public information on upcoming trades in a market.
In the context of blockchain, front-running a transaction refers to submitting a transaction which is solely intended to be executed \textit{before} some other pending transaction~\cite{eskandari2019sok}. 
As transactions are executed sequentially according to how they have been ordered in a block, an agent may financially benefit from front-running one or more transactions, by having their transaction executed before a victim transaction.
Similarly, an agent may pursue \textit{back-running}, whereby a transaction is intended to be executed \textit{after} some designated transaction.
As the majority of Ethereum miners order transactions by their gas price~\cite{zhou2020high}, an agent can set a higher or lower gas price relative to some target transaction, in order to have his transaction executed before or after the target, respectively.
In the case of multiple agents attempting to front-run the same transaction, front-running results in priority gas auctions~\cite{daian2019flash}, i.e. the competitive bidding of transaction fees to obtain execution priority.

We refer to attacks which involve front- and/or back-running within a single block, thereby undermining the technical security of DeFi protocols, as transaction ordering attacks.
Note that an attacker does not need to be a miner in order to execute the following attacks but such attacks can be undertaken risk-free if the attacker is a miner.

\point{Displacement Attacks}
In a displacement attack, an attacker front-runs some target transaction, where the success of the attack does not depend on whether the target transaction is executed afterwards~\cite{eskandari2019sok}.
A simple example of such an attack would be an attacker front-running a transaction that registers a domain name~\cite{kalodner2015empirical}.
A more severe risk comes in the form of \textit{generalized} front-runners~\cite{robinson2020ethereum}, which are bots that parse all unconfirmed transactions in the mempool, trying to identify, duplicate, modify and lastly front-run any transaction which would result in a financial profit to the front-runner.
Examples of transactions vulnerable to generalized front-runners would be reporting a bug as part of a bug bounty scheme to claim a reward~\cite{breidenbach2018enter} and trying to `rescue' funds from an exploitable smart contract~\cite{robinson2020ethereum,samczsun2020escaping}.

% A further example would be when a sender intends to to make a risk-free profit within a single transaction, it can be vulnerable to displacement attacks by \textit{generalized} front-runners~\cite{robinson2020ethereum}. 
% These bots parse all unconfirmed transactions in the mempool, trying to identify, duplicate, modify and lastly front-run any transaction which would result in a financial profit to the front-runner.

% A further vector for displacement attacks applies to order book DEXs, on which exchange participants are required to submit transactions to cancel existing orders.
% If a user submits a transaction to cancel an unfilled order due to price changes before the order could be filled, an attacker could front-run the cancel transaction and fill the order.
% In the context of DEXs, the success of such front-running behavior is particularly likely given the widespread existence of arbitrage bots engaging in PGAs for execution priority~\cite{daian2019flash}.

\point{Multi-transaction Sandwich Attacks}
In a ``sandwich attack", an attacker alters the deterministic price on an AMM prior to and after some other target transaction has been executed in order to profit from temporary imbalances in the AMM's liquidity reserves.
In simple cases (e.g., Uniswap), the instantaneous AMM price is simply a ratio of AMM reserves and imbalances can be created simply by changing this ratio (e.g., by providing single-sided liquidity or performing a large swap through the AMM). 
This is how these AMMs are designed to work: swaps create imbalances, which, if left unbalanced, incentivize arbitrageurs to perform the reverse actions to balance the AMM pool.

An attacker may target another user's transaction (e.g., to profit from triggering large slippage in another user's swap) by trying to place adjacent transactions that set up the imbalance right before the swap and close out the imbalance right after the swap~\cite{zhou2020high,swende2017blockchain}.
This can be achieved through front-running the user's swap transaction by setting a higher gas price on the transaction creating the imbalance.
By setting a lower gas price on the transaction closing the imbalance, the attacker can back-run the user's transaction and complete the attack.
Note that setting high and low transaction fees does not guarantee the attack to succeed, as ultimately it is up to a transaction's miner to determine the order of execution.

A variant of this attack~\cite{zhou2020high} can be performed if instead of being a liquidity taker, the attacker is a liquidity provider for the respective AMM.
The attacker can front-run a victim transaction that swaps token $A$ for token $B$ and remove liquidity, exposing the victim to higher slippage.
Subsequently, the attacker can back-run the victim transaction, and resupply the previously withdrawn liquidity.
In a third transaction that swaps $B$ for $A$, the attacker obtains a profit in $B$.
A formal analysis of sandwich attacks is given in \cite{zhou2020high}.

\hidelinks
\begin{table}[t]
  \setlength{\tabcolsep}{2pt}
  \centering
  \begin{tabular}{l r r l l r}
    \toprule
        \thead[l]{\textbf{Protocol}} & \thead[c]{\bf Loss} & \thead[c]{\bf Audit} & \thead[l]{\textbf{Attack}} & \thead[l]{\textbf{Date}} & \thead[c]{\textbf{Ref.}}\\
    \midrule
    bZx & $0.35$m & \cmark & TX sandwich & Feb-$15$-$2020$ & \cite{hack:bzx3}\\
    bZx & $0.63$m & \cmark & TX sandwich & Feb-$18$-$2020$  & \cite{hack:bzx2}\\
    Uniswap & $0.30$m & \cmark & Reentrancy & Apr-$18$-$2020$ & \cite{hack:imbtc-uniswap} \\
    dForce & $25.00$m & \xmark & Reentrancy & Apr-$19$-$2020$ & \cite{web:dforce-hack}\\
    Hegic & $0.05$m & \xmark & Logical bug & Apr-$25$-$2020$ & \cite{hack:hegic} \\
    Balancer & $0.50$m & \cmark & TX sandwich & Jun-$28$-$2020$ & \cite{hack:balancer} \\
    Opyn & $0.37$m & \cmark & Logical bug & Aug-$04$-$2020$ & \cite{hack:opyn}\\
    Yam & $0.75$m & \xmark & Logical bug  & Aug-$12$-$2020$  & \cite{web:yam-bug}\\
    bZx & $8.10$m & \cmark & Logical bug & Sep-$14$-$2020$ & \cite{hack:bzx}\\
    Eminence & $15.00$m & \xmark & TX sandwich & Sep-$29$-$2020$ & \cite{hack:eminence}\\
    MakerDAO & - & \cmark & Governance & Oct-$26$-$2020$ & \cite{makerdao2020flash} \\
    Harvest & $33.80$m & \cmark &  TX sandwich & Oct-$26$-$2020$ & \cite{hack:harvest}\\
    Percent & $0.97$m & \cmark & Logical bug & Nov-$04$-$2020$ & \cite{hack:percent-finance} \\
    Cheese Bank & $3.3$m & \cmark & TX sandwich & Nov-$06$-$2020$ & \cite{hack:cheese}\\
    Akropolis & $2.00$m & \cmark & Reentrancy & Nov-$12$-$2020$ & \cite{hack:akropolis}\\
    Value DeFi & $7.00$m & \xmark & TX sandwich & Nov-$14$-$2020$ & \cite{peckshield2020value}\\
    Origin & $7.00$m & \cmark & Reentrancy & Nov-$17$-$2020$  & \cite{hack:origin}\\
    88mph & $0.01$m & \cmark & Logical bug & Nov-$17$-$2020$ & \cite{hack:88mph}\\
    % Bancor & \$$0.14$m & & \xmark & Displacement attack & Jun-$17$-$2020$ \\
    Pickle & $19.70$m & \xmark & Logical bug  & Nov-$21$-$2020$ & \cite{hack:pickle-finance} \\
    % Soft Yearn & \$$0.25$m & \xmark & [uniswap price not adjusted in time after rebase] & Sep-$03$-$2020$ \\
    Compounder & $10.80$m & \cmark & Logical bug & Dec-$02$-$2020$ & \cite{hack:compounder} \\
    Warp Finance & $7.80$m & \cmark & TX sandwich & Dec-$18$-$2020$ & \cite{rekt2020warp} \\
    Cover & $9.40$m & \cmark & Logical bug & Dec-$28$-$2020$ & \cite{hack:cover}\\
    % Saddle Finance & $0.28$m & \cmark & & Jan-$20$-$2021$ & \cite{rekt2021saddle}\\
    Yearn & $11.00$m & \xmark & TX sandwich & Feb-$05$-$2021$ & \cite{rekt2021yearn}\\
    Growth DeFi & $1.30$m & \cmark & Logical bug & Feb-$09$-$2021$ & \cite{rekt2021growth}\\
    Meerkat & $32.00$m & \xmark & Logical bug & Mar-$04$-$2021$ & \cite{rekt2021meerkat}\\
    Paid Network & $27.00$m & \xmark & Logical bug & Mar-$05$-$2021$ & \cite{rekt2021paid}\\
    DODO & $2.00$m & \xmark & Logical bug & Mar-$09$-$2021$ & \cite{rekt2021dodo}\\
    Cream & $130.00$m & \cmark & TX sandwich & Oct-27-2021 & \cite{yearn-cream2021}\\
    % Roll & $5.70$m & \xmark & & Mar-$14$-$2021$ & \cite{rekt2021roll}\\

    \bottomrule
  \end{tabular}
  \caption{An overview of empirical technical security exploits in DeFi protocols for the period February 2020 to March 2021. The included exploits are explicitly limited to technical exploits and exclude any deliberate protocol scams that may have occurred. 
  Note that the amount of funds seized per exploit is denominated in USD as of the time of the exploit and does not account for any losses that may have been recovered.}
  \label{tab:technical-attacks}
\end{table}
\restorelinks

%% file: content/5_economic_security.tex
\section{Economic security}
\label{sec:economic-security}

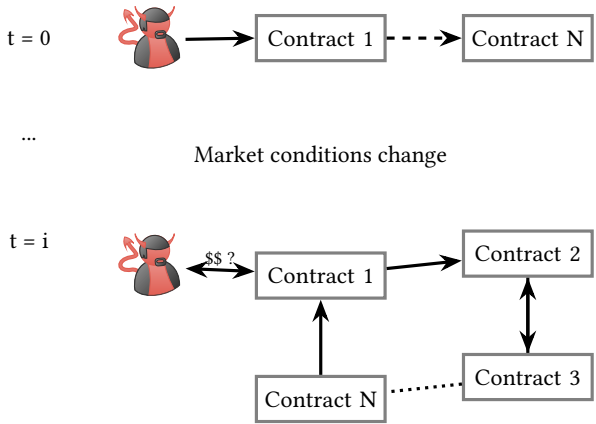
\begin{figure}[htp]
    \centering
        \begin{tikzpicture}[auto,
                       > = Stealth, 
           node distance = 10mm,
              box/.style = {draw=gray, very thick,
                            minimum height=6mm, text width=15mm, 
                            align=center},
       every edge/.style = {draw, very thick},
  every edge quotes/.style = {font=\footnotesize, align=center, inner sep=1pt}
                            ]
  % from bottom to top
    \node (n11) [] {t = 0};
    \node (n12) [devil,evil,minimum size=6mm, right=of n11] {};
    \node (n13) [box, right=of n12] {Contract 1};
    \node (n14) [box, right=of n13] {Contract N};
    \node (n21) [below=of n11] {...};
    \node (n22) [below=of n12] {};
    \node (n23) [below=of n13] {Market conditions change};
    \node (n24) [below=of n14] {};
    \node (n31) [below=of n21] {t = i};
    \node (n32) [devil,evil,minimum size=6mm, below=of n22] {};
    \node (n33) [box, below=of n23] {Contract 1};
    \node (n34) [box, below=of n24] {Contract 2};
    \node (n41) [below=of n31] {};
    \node (n42) [below=of n32] {};
    \node (n43) [box, below=of n33] {Contract N};
    \node (n44) [box, below=of n34] {Contract 3};
  %Lines
  \draw
        (n12) edge [->]     (n13)
        (n13) edge [dashed, ->] (n14)
        (n32) edge [<->, "\$\$ ?"]                   (n33)
        (n33) edge [->]                   (n34)
        (n44) edge [->]               (n34)
        (n34) edge [->]                   (n44)
        (n44) edge [dotted]                   (n43)
        (n43) edge [->]       (n33);      
        \end{tikzpicture}
        \caption{Diagram of an economic exploit.}
        \label{fig:economic-security-illustration2}
  \end{figure}

We define a DeFi security risk to be \emph{economic} if an attacker can perform a strictly non-atomic exploit to realize a profit at the expense of value held by the protocol or its users.
In an economic exploit, an attacker performs multiple actions at different places in the transaction sequence but doesn't control what happens between their actions in the sequence, which means that there is no guarantee that the final action is profitable (as visualized in Fig~\ref{fig:economic-security-illustration2} and in comparison to the technical exploit in Fig~\ref{fig:technical-security-illustration}).
Economic security is effectively about an exploiting agent who tries to manipulate a market or incentive structure over some time period (even if short, it is not instantaneous). Compared to technical exploits, since economic exploits are non-atomic, they come with upfront tangible costs, a probability of attack failure and risk related to mis-estimating the market response. Thus they are not risk-free and commonly involve manipulations over many transactions or blocks.

In addition to comparing the structures in Figures \ref{fig:technical-security-illustration} and \ref{fig:economic-security-illustration2}, we provide a simple example to help illustrate the distinction between technical and economic security. Consider a protocol that uses an instantaneous AMM price as an oracle. An attacker can perform a (atomic) sandwich attack to steal assets, which amounts to a technical exploit. If instead the protocol used a time-weighted average AMM price as an oracle, then the attacker could manipulate this price over time (non-atomically) and may still be able to steal assets, which would amount to an economic exploit.

Economic risks are inherently a problem of economic design and cannot be solved by technical means alone. To illustrate, while these attacks could be performed atomically (and risk-free) in a very poorly constructed system that allowed it, they are not solved, for example, just by adding a time delay that ensures they are not executed in the same block. Even if all technical issues are sorted, we are often left with remaining economic problems about how markets or other incentive structures could be manipulated over time to exploit protocols.
From a practical perspective, progress on these economic problems inherently requires economic models of these market equilibria and the design of better protocol incentive structures. These models differ considerably from traditional security models and are sparsely studied. As a result, defensive measures for economic security risks are also not as well established.

In this way also, technical security must be a first bar: if a protocol is not technically secure, then it will break in the presence of rational agents. Economic security only makes sense if technical security is achieved. For instance, if a protocol's funds can be exploited because it is not technically secure, then in an economic sense no rational agents should participate.

\begin{tcolorbox}[boxsep=1pt,left=2pt,right=2pt,top=2pt,bottom=2pt, title=Economic Security]
\emph{A DeFi protocol is economically secure if it is economically infeasible (e.g., unprofitable) for an attacker to perform exploits that are strictly non-atomic at the expense of value held by the protocol or its users. As economic exploits are non-atomic (or else they are better described as technical), they are not risk-free.}
\end{tcolorbox}

%%%%%%%%%%%%%%%%%%%%%%%%%%%%%%%
\point{Economic Rationality}
\label{sec:economic-rationality}
A central assumption in considering the class of economic security attacks is that of economic rationality.
Following the standard game theoretic approach, we denote the strategy for player $i$ as $s_i$.
A strategy is a plan for what to do at each decision node (equivalently, information set) that the agent is aware they might reach.
For example,  a strategy would define what action an agent would take in the event that it finds itself in a protocol that becomes undercollateralized.
A strategy $s_{1,i} \in \S_i$ for player $i$ strictly dominates another strategy $s_{2,i} \in \S_i$ if regardless of the actions of other agents, strategy $s_{1,i}$ will always result in a higher payoff to the agent. 
Economic rationality is then defined as follows. 

\begin{tcolorbox}[boxsep=1pt,left=2pt,right=2pt,top=2pt,bottom=2pt, title=Economic Rationality]
\emph{An agent is rational iff they will never play a strictly dominated strategy (i.e., they are profit optimizing).} 
\end{tcolorbox}

Moreover, common knowledge of rationality means that all agents know no agent will play a strictly dominated strategy.

While most economic security analysis ought to consider attackers who have profit-maximizing objectives, it can also be important to consider attackers with other objectives. For instance, an attacker who wishes to shut down the system may decide to attack as long as the cost is of a moderate level. 
In this sense, the economic security depends on system interruptions being too costly to effect.

%%%%%%%%%%%%%%%%%%%%%%%%%%%%%%%
\point{Incentive Compatibility}
\label{sec:incentive-compatibility}
\textit{Incentive compatibility} is originally a concept from game theory (e.g., ~\cite{roughgarden2010algorithmic}, but as a concept has seen some adaption in the context of cryptoeconomics and in particular DeFi. 

In the cryptoeconomic setting, a mechanism is defined as incentive compatible if agents are incentivized to execute the mechanism \textit{as intended} (see e.g.~\cite{roughgarden2020transaction}).

\begin{tcolorbox}[boxsep=1pt,left=2pt,right=2pt,top=2pt,bottom=2pt, title=Cryptoeconomic Incentive Compatibility]
A mechanism (or protocol) is incentive compatible iff agents are incentivized to execute the game as intended by the protocol designer.
\end{tcolorbox}

A central question in the context of incentive compatibility, considered in \cite{klages2020stablecoins}, is the sustainability of the mechanism implemented by a system (i.e., will the incentives arising from the system allow the system to be economically secure and stable long-term).
In \cite{klages2020stablecoins}, for stablecoins, this is separated into a question of incentive security, which is included in our concept of economic security, and a question of economic stability, which is a further question of whether an economically secure system actually plays out to the desired equilibrium envisioned by the designers.

We primarily focus on the direct security questions in this paper; however, similar questions to economic stability apply to protocols other than stablecoins as well.
For instance, when designing synthetic derivatives built using dynamic portfolios (and implemented as AMM pools), a lingering question is how well these designs can replicate the derivative payoffs under extreme conditions. 
As a comparison, synthetic portfolio insurance in traditional markets can break down when markets move too fast for the strategy to rebalance (See Ch. 13 in \cite{hull2009options}). 
AMM pools aim to rebalance over much shorter timescales, and so may have an advantage here, but are also suboptimal in other areas of rebalancing.

%%%%%%%%%%%%%%%%%%%%%%%%%%%%%%%%%%%%%%%%%%%%%%%%
\subsection{Overcollateralization as Security}
Collateralization is one of the primary devices to ensure economic security in a protocol.
In general, collateral serves as a potential repercussion against misbehaving agents~\cite{harz2019balance} and allows creating protocols such as stablecoins, loanable funds, or decentralized cross-chain protocols.
% As outlined in Section~\ref{sec:plfs}, in a trustless system without strong identities or legal recourse, overcollateralization creates the economic incentive for the loan to be repaid, or at least insures the lender against losses. 
As asset prices evolve over time, these systems generally allow automated deleveraging: if an agent's level of collateralization (value of collateral / value of borrowing) falls below a protocol-defined threshold, an arbitrager in the system can reduce the agent's borrowing exposure in return for a portion of their collateral at a discounted valuation. This aims to keep the system fully collateralized.

Overcollateralization is not without risks, however. For instance, as explored in \cite{gudgeon2020decentralized,kao2020analysis}, times of financial crisis (wherein there are persistent negative shocks to collateral asset prices) can result in thin, illiquid markets, in which loans may become undercollateralized despite an automated deleveraging process. 
In such settings, it can become unprofitable for liquidators, a type of keeper, to initiate liquidations. 
Should this occur, rational agents will leave their debt unpaid as that results in a greater payoff.

Another form of deleveraging risk arises when the borrowed asset has endogenous price effects, for instance when its price is affected by other agents' decisions in the system or when it is manipulable. 
This is the case in non-custodial stablecoins like Dai that are based on leverage markets (Dai is created by ``borrowing'' it against collateral and similarly must be returned to later release the collateral). 
As explored in \cite{klages2019stability,klagesmundt2020while}, such stablecoins can have deleveraging feedback effects that lead to volatility in the stablecoin itself. 
In regions of instability, the stablecoin will tend to become illiquid and appreciate in price (more so as they need to be purchased for liquidations), which can force speculative agents who have leveraged their positions to pay premium prices to deleverage. 
This causes their collateral to drawdown faster than may be expected, which makes the system in total less healthy and may lead to shortfalls in collateralization. 
This was later directly observed in Dai on ``Black Thursday''~\cite{frangella2020crypto}. 
As further discussed in \cite{klagesmundt2020while}, such a stablecoin requires uncorrelated collateral assets to be fully stabilized from such deleveraging effects as stable regions are related to submartingales (i.e., agents expect collateral asset prices to appreciate). 
However, current uncorrelated assets are primarily centralized/custodial, which poses a challenge for non-custodial designs.

\subsection{Threats from Miner Extractable Value}
\label{sec:mev}
An assumption by many blockchain protocols is that the block reward is sufficient to incentivize ``correct" miner behavior.
However, there are consensus layer risks should the MEV exceed the block reward.
The simplest example of MEV is double spending of coins, which is commonly considered in base layer incentives. 
DeFi applications give rise to many new sources of MEV. 
For instance, (1) DEXs present atomic arbitrage opportunities between different trading pairs, as explored in \cite{daian2019flash}, and (2) stablecoins built on leverage markets (like Dai) present arbitrage opportunities in liquidating leveraged positions, as explored in \cite{klages2019stability}.
Similarly, other protocols, like PLFs, that utilize liquidation mechanisms also create MEV opportunities. 
Further, MEV can arise when miners are incentivized to re-order or exclude transactions based on cross-chain payments happening on other chains \cite{judmayer:p2w}.

These are not exhaustive; there are additionally many other ways in which miners could manipulate DeFi protocols to extract value. 
It's worth noting that these are not just hypothetical concerns, they have actually been observed--e.g.,~\cite{blocknative2020,pegnet2020}--and shown to be sufficiently profitable, e.g.,~\cite{zhou2021just}.

The practicality of MEV threats have been highlighted in~\cite{daian2019flash}, where the prevalent dangers of \textit{undercutting} and \textit{time-bandit} attacks are presented.
In an undercutting attack~\cite{carlsten2016instability}, an adversarial miner would fork off a block with high MEV, while holding back some of the extractable value in order to incentivize other miners to direct their computational efforts towards the adversary's chain.
In a time-bandit attack~\cite{daian2019flash}, an attacker forks from some previous block and sources \textit{expected} MEV to increase his computational power and pursue a $51$\% attack until the expected MEV is realized. 
Hence, time-bandit attacks are a consensus layer risk and can be a direct consequence of historic on-chain actions which could profit a miner at some later point. A further threat is that miners could collude to set up more MEV opportunities over time, for instance by censoring transactions to top up collateral in crises and thus creating more liquidation events, as discussed in~\cite{klages2019stability}. This is very similar to events on Black Thursday, in which mempool manipulations contributed to inefficient liquidation auctions in Maker~\cite{blocknative2020}.

%%%%%%%%%%%%%%%%%%%%%%%%%%%%%%%%%%%%%%%%%%%%%%%
\subsection{Governance Risks and Governance Extractable Value}

Protocol governance often introduces means to update system parameters and even redefine entire contracts. 
In many cases, this may be a necessary component for the system to evolve over time. However, governance can also introduce manipulation vectors that affect security. 
Governance of a DeFi protocol is typically tied to holders of governance tokens, which can often be thought of as shares in the protocol.
In systems where there is large flexibility for governance to change the system, an important question is where governance token value comes from.
A typical aim is for the protocol to incentivize good stewardship from its governance token holders by compensating governance with cashflows from the system. 
In this case, governance token value is derived from future discounted cashflows. 
Another possibility is that governance is directly aligned with underlying users--e.g., because they are the same.

However, if these incentives are not of sufficient size, then it may be more profitable for governance token holders to extract value in less desirable ways, which we term \emph{governance extractable value} (GEV).\footnote{GEV may be interpreted as a generalization of MEV with miners being a specific type of governor tasked with ordering transactions on the base layer.} 
An example of GEV is for governors to effect changes to the protocol in ways that provide outside benefits to themselves but may be harmful to the overall system health. 
For instance, the Cream protocol governance added high levels of very risky collateral assets that they had an outside stake in, arguably to their benefit but against the interests of the protocol \cite{cream_ftt}.

GEV also includes explicit governance attacks.
An instance of an explicit attack was the governance takeover of the Build Finance DAO, where a malicious actor passed a proposal to take control of the Build token contract and was thereby not only able to drain various AMM pools by minting and swapping Build tokens, but to ultimately remove the DAO from any form of control over the core protocol~\cite{build_finance_gev}.
A hypothetical GEV attack to indirectly extract collateral value is described in \cite{klagesmundt2019vuln}.
In cases like these, governance may not be incentive compatible. 
And if the value of governance tokens from incentive compatible sources crashes, the region of incentive compatibility also shrinks, and it may become profitable for a new coalition of governors to form to attack the protocol. 
This is increasingly problematic given the ease and low cost with which governance tokens may be obtained via flash loans and PLFs. 
Other complications arise in the need to protect minority rights within the protocol--e.g., building in limitations so that a majority of governors cannot unilaterally change the game to, for instance, steal all value of the other minority or users. See \cite{lee2021gov} for further discussion of GEV.

There is limited literature in modeling GEV incentives in the DeFi setting (as opposed to modeling governance in the underlying blockchain itself). 
% A starting point is introduced in \cite{klages2020stablecoins} as a framework of capital structure-like models to model GEV and incentive compatibility.
The capital structure-like models developed in \cite{klages2020stablecoins, huo2022decentralized} can be applied more generally to DeFi protocols to model governance security and incentive compatibility around these issues. 
% Compared to (idealized) traditional financial settings, DeFi exists in a pseudo-anonymous setting that lacks strong outside recourse mechanisms (e.g., legal) to disincentivize attacks and manipulations.
As can be understood in those models, these issues essentially arise because there may not be outside recourse (e.g., legal) in the pseudo-anonymous setting to disincentivize attacks and manipulations compared to the (idealized) traditional finance setup. Further, \cite{klages2020stablecoins} conjectures that in the case of a fully decentralized stablecoin with multiple classes of interested parties and with a high degree of flexibility for governance design, there exists no long-term incentive compatible equilibrium. Intuitively, there are resulting costs of anarchy in such systems, which can be too much to bear. In such a case, rational agents would choose not to participate. However, they also conjecture that other DeFi systems, such as DEXs, may have wider incentive compatibility in similar situations due to the different structure of such systems.
% There are resulting costs of anarchy in such systems, that can drive down the incentives to participate in truly decentralized systems.
These models have inspired new mechanisms such as Optimistic Approval \cite{lee2021gov}, which provides an optional veto over governance updates to protocol users, as a new defensive measure to to reduce costs of anarchy and GEV.

% The capital structure-like models developed in \cite{klages2020stablecoins} can be applied more generally to DeFi protocols to model governance security and incentive compatibility around these issues. As can be understood in those models, these issues essentially arise because there may not be outside recourse (e.g., legal) in the pseudo-anonymous setting to disincentivize attacks and manipulations compared to the (idealized) traditional finance setup. Further, \cite{klages2020stablecoins} conjectures that in the case of a fully decentralized stablecoin with multiple classes of interested parties and with a high degree of flexibility for governance design, there exists no long-term incentive compatible equilibrium. Intuitively, there are resulting costs of anarchy in such systems, which can be too much to bear. In such a case, rational agents would choose not to participate. However, they also conjecture that other DeFi systems, such as DEXs, may have wider incentive compatibility in similar situations due to the different structure of such systems.

%%%%%%%%%%%%%%%%%%%%%%%%%%%%%%%%%%%%%%%%%%%%%%%
\subsection{Market and Oracle Manipulation}
\label{sec:market-manipulation}
As the suppliers of off-chain information, oracles pose a fundamental component of DeFi protocols, particularly for sourcing price feeds \cite{kaleem2021demystifying}.
However, it is important to distinguish between (1) a price that is manipulated yet correctly supplied by an oracle and (2) an oracle itself being manipulated. 
While we present each form of manipulation, note that the latter can be essentially modeled as a separate governance-type risk as discussed in \cite{klages2020stablecoins}.

\point{Market Manipulation}
We wish to quantify economic risks stemming from price manipulations in underlying markets while assuming the oracle follows a best practice implementation and is non-malicious.
An adversary may manipulate the market price (on-chain or off-chain) of an asset over a certain time period if a profit can be realized as a consequence of the price manipulation--e.g., by taking positions in a DeFi protocol that uses that market price as an oracle.
As discussed in the Section~\ref{sec:technical-security}, instantaneous AMM prices are easily manipulable with near zero cost and, as a result, should not be used as price oracles.
Market manipulation problems persist even when we assume the oracle is not an instantaneous AMM price. 
In this case, there is a cost to market manipulation related to maintaining a market imbalance over time, whether in an AMM (e.g., to manipulate a time-weighted average price) or through filling unfilled orders in an order book. 
Depending on whether the market for an asset is thick or thin, the cost for an attacker to significantly change the asset's price will be higher or lower, respectively. 
An instance of this form of market manipulation was seen with Inverse Finance, where a malicious actor first manipulated the used Sushiswap price oracle to quote a higher price for the Inverse token, only to exploit the protocol a block later by borrowing various assets against the Inverse token using the manipulated price before MEV bots arbitraged the manipulated price back~\cite{inverse_exploit}.
A further example of such an attack would be to trigger liquidations by manipulating an asset's price, as discussed in the context of stablecoins in \cite{klages2019stability}.
An attacker could profit either by purchasing liquidated collateral at a discount or shorting the collateral asset by speculating on a liquidation spiral.
Such attacks are similar to short-squeezes in traditional markets.
However, unlike with single transaction sandwich attacks, the aforementioned attack is not risk-free and could bring substantial losses to the attacker should it fail. 
In particular, markets and agents may react to such attacks in unpredictable ways.
% Two examples to illustrate such exploits are given in Appendix~\ref{app:empirical-exploits:market-manip}.

To  illustrate the potential of such attacks, the stablecoin DAI, which historically has thin liquidity, traded at a temporary price of \$$1.30$ over a course of about $20$ minutes on Coinbase Pro, a major centralized cryptoasset exchange, before returning to its intended \$$1$ peg~\cite{khatri20dai}. 
As a result, the Compound Open Price Feed~\cite{web:open-price-feed}, a cryptoasset price oracle which is in part based on prices signed by Coinbase, reported a DAI price of \$$1.23$ to Compound for a short period of time. 
This incident triggered (arguably wrongful) liquidations on collateral worth approximately \$$89$m, costing the liquidated Compound borrowers $23$\% (from the imbalanced DAI price) plus an additional $5$\% (the Compound liquidation incentive, i.e., the discount at which collateral is sold at during a liquidation) on their liquidated assets.

\point{Oracle Manipulation}
Centralized oracles serve as a single point of failure and despite trusted execution environments~\cite{zhang2016town} they remain vulnerable to the provider behaving maliciously if incentives are sufficient for manipulating the source of a data feed. 
Decentralized price oracles may use on-chain data, most notably on DEXs (specifically AMMs) for crypto-to-crypto price data.
However, as outlined in Section~\ref{sec:technical-sec:single-tx-sandwich}, prices may be manipulable through intentionally created imbalances and thinly traded markets.
%, even after remedying the technical security issues using, for instance, time-weighted average prices.
Furthermore, on-chain DEX oracles inherently can not price off-chain assets and fiat currencies. 
%For instance, cryptoasset prices may be quoted in stablecoins through DEX oracles, but this faces the same inherent problem: we then rely on that stablecoin, which may be manipulated or fail, for the data feed.

As discussed in \cite{klages2020stablecoins}, decentralized oracle solutions for off-chain data exist. 
However, they are yet imperfect solutions, tending to rely on Schelling point games, in which agents vote on the correct price values and are incentivized against having their stake slashed if their vote deviates from the consensus.
Tying incentives to consensus, when the correctness of the consensus decision is not objectively verifiable (as in this case), paves a vector for game theoretic attacks, like in Keynesian beauty contests.
% Widely used decentralized oracles, such as Chainlink~\cite{ellis2017decentralized}, try to mitigate this problem by aggregating data feeds from multiple sources (e.g., by calculating the median) and relying on reputation systems to curate reliable sources.
% These systems may still suffer from similar game theoretic issues, however.

%% file: content/6_challenges.tex
\section{Open Research Challenges}
\label{sec:challenges}

There are many open research challenges in DeFi stemming from the technical and economic security issues presented in Sections \ref{sec:technical-security} and \ref{sec:economic-security}.

\subsection{Composability Risks}

Cryptoassets can be easily and repeatedly tokenized and interchanged between DeFi protocols in a manner akin to rehypothecation. 
This offers the potential to construct complex, inter-connected financial systems, yet bears the danger of exposing agents to composability risks, which are as of yet mostly unquantified.
An example of composability risk is the use of flash loans for manipulating instantaneous AMMs and financially exploiting protocols that use those AMMs as price feeds. 
This has repeatedly been exploited in past attacks (e.g.~\cite{hack:harvest,hack:value,hack:cheese}). 
% Many protocols still struggle to implement sufficient protective measures for addressing this risk.
However, the breadth of composability risks spans far beyond the negative externalities stemming from instantaneous AMM manipulations.
For instance, there remain open questions about the consequences of the following types of exploitations on connecting systems: the accumulation of governance tokens to execute malicious protocol updates, the failure of non-custodial stablecoin incentives to ensure price stability, and failure of PLF systems to remain solvent. 
Note, however, that this list is far from exhaustive.
These become increasingly important issues as more complex token wrapping structures \cite{von2021measuring} stimulate higher degrees of protocol interconnectedness. 
For example, the use of PLF deposit tokens (as opposed to the tokens in their original forms) within AMM pools and strategies to earn yield on underlying assets through leverage by borrowing non-custodial stablecoins and depositing into PLFs or AMMs.

Recent works \cite{gudgeon2020defi,nadler2020decentralized} begin to explore protocol interdependence, while \cite{tolmach2021formal} propose a process-algebraic technique
that allows for property verification by modeling DeFi protocols in a compositional manner. Nonetheless, a critical gap in DeFi research toward taxonimizing and formalizing models to quantify composability risks remains.
This problem is elevated as a holistic view on the integrated protocols is necessary: failures might arise from both technical and economic risks.
Ensuring safety of protocol composition will be close to impossible for any protocol designer and forms a major challenge for DeFi going forward.

\subsection{Governance}

We identify important research directions in governance and GEV.
A general direction is modeling incentive compatibility of governance in various systems with GEV.
For instance, setting up models, finding equilibria, and understanding how other agents in the system respond. 
The models in \cite{klages2020stablecoins} get this started in the context of stablecoins and additionally discuss how to extend to other DeFi protocols. 
There is moreover a range of discussions around simulating and formalizing governance incentives through tools like cadCAD~\cite{web:cadcad}.

There remain unanswered questions with regards to the general design of governance incentives. 
For instance, how to structure governance incentives to reward good stewardship: e.g., intrinsic vs. monetary reward, reward per vote vs. reward per token holder, and measures of good stewardship.
Furthermore, there lies potential in formally evaluating protection of minority agents in systems with flexible governance and large GEV.

For systems utilizing governance tokens, we identify research gaps rooted in security risks of the ability to borrow governance tokens via flash loans and PLFs.
Specifically, we identify opportunities to formally explore how (1) technical security can be compromised and (2) from an economic security point of view, incentive compatibility is further complicated by the borrowing of governance tokens.

\subsection{Oracles}
\label{sec:oracle-challenges}

We highlight a few open challenges about oracle design and security. 
Note that, in many cases, the oracle problem can also be directly related to the governance problem, as governors are often tasked with choosing the oracles that are used.

A more general open challenge lies in how to structure oracle incentives to maintain incentive compatibility to report correct prices.
This is similar to governance design in some ways and needs to take into account the possible game theoretic manipulations that could be profitable.

We identify a further research opportunity in designing and evaluating the security of various oracle strengthening methods.
While there exist several works examining oracle designs on both a general and empirical basis--e.g. \cite{liu2020look,kaleem2021demystifying}-- a formal security analysis of, e.g., medianizers, reputation systems, and grounding reported prices based on on-chain verifiable metrics is yet to be done.

\subsection{Miner Extractable Value}
% We identify important research directions in MEV.

While research on MEV and the extraction of it is being put forward~\cite{daian2019flash}, methodologies to quantify negative externalities of MEV--e.g., from wasted gas per block, upward gas price pressure--and the full extent of MEV opportunities remain scarce.
For the latter, we conjecture that the miner's problem to optimize the MEV they extract in a block is NP-hard and additionally hard to approximate. 
To support this, it is quite easy to reduce a simplified version of the problem, in which the MEV of each transaction is fixed, to the knapsack problem. 
Note that while the knapsack problem is NP-hard, it is easy to approximate. 
In fact, we expect a more realistic version of the miner's problem to be harder than knapsack because the transaction ordering the miner chooses also changes the MEV of the transactions (i.e., swapping two elements might change their weight in knapsack).

With regards to quantifying extractable value, \cite{babel2021clockwork} introduce Clockwork Finance Framework, a novel framework that applies formal verification to reason about DeFi security properties with respect to the profit that is extractable by an actor in a given system.
It should be noted that the definition of economic security as defined in~\cite{babel2021clockwork} would be closer to what we refer to as technical security or atomic MEV and that modeling non-atomic MEV requires economic models of the underlying markets, as  discussed in Section~\ref{sec:economic-security}.

There are interesting questions regarding how the emergence of MEV opportunities endogenously affects agents' behavior within DeFi protocols. 
Models for this are started in the context of stablecoins in~\cite{klages2020stablecoins}.

A further open challenge remains with respect to designing protective mechanisms against (1) consensus layer instability risks that are induced by high MEV incentives and (2) time bandit attacks that seek to rewrite the recent transaction history--for example, which could aim to trigger and profit from increased protocol liquidations.
On this point,~\cite{klages2019stability} suggests that oracle price validity could be tied to recent block hashes to prevent such reorderings from extracting the protocol value, though potentially with costs to the economic security of the protocol in other ways.

\subsection{Program Analysis}
There exists a large amount of work~\cite{harz2018towards}, both in academia~\cite{luu2016making,tsankov2018securify,permenev2020verx} and industry~\cite{web:mythx,web:slither}, to analyze smart contract bugs and vulnerabilities.
While smart contracts analysis tools keep improving, the number and scale of smart contract exploits are showing no sign of decrease and are, on the contrary, becoming more frequent.
Although program analysis tools are no silver bullet and cannot prevent all exploits, Table~\ref{tab:technical-attacks} and the discussed exploits in Section~\ref{sec:technical-security} and in Appendix~\ref{app:empirical-exploits} hint that there are some recurring patterns that could be automatically detected and prevented.
% However, recent exploits (as shown in Table~\ref{tab:technical-attacks} and Appendix~\ref{app:empirical-exploits}) hint that there are some recurring patterns that could be automatically detected and prevented.
% We argue that improvements in program analysis could prevent many of the exploits we have seen.

Current program analysis tools can mainly be divided into two categories: (1) fully automatic tools checking for program invariants and (2) semi-automated verification tools checking for user-defined properties~\cite{permenev2020verx,annenkov2019towards,chen2018language}.
While the latter allows to verify business logic in ways that are not fully automatable, they are typically non-trivial to setup and require knowledge of software verification, which limits their use to projects with enough resources.
On the other hand, fully automatic tools, which can be very easily setup and ran, usually focus on checking properties of a single contract in isolation~\cite{tsankov2018securify,10.1145/3274694.3274737,web:slither,web:mythril}, such as unchecked exceptions or integer overflows.
However, they have not evolved yet to embrace the composable nature of smart contracts, which makes it impossible for such tools to reason about scenarios where the issue happens due to a change in something external to the smart contracts, such as a sudden change in a price returned by an oracle.
Further, most tools reason very little about \emph{semantic} properties of the smart contracts, such as how a particular execution path can influence ERC-20 token balances.
We believe that improvements in these areas will allow auditors and developers to analyze and deploy their contracts with more confidence, reducing the number of technical security exploits.

\subsection{Anonymity and Privacy}
The anonymity and privacy of DeFi protocols is at present a significantly understudied area.
There is a tension between user's privacy being valuable in itself, while at the same time helping malicious users to escape the consequences of their actions.
At present, a large proportion of DeFi transactions occurs in protocols built on Ethereum, wherein agents at best have pseudoanonymity.
This means that if an agent's real-world identity can be linked to an on-chain address, all the actions undertaken by the agent through that address are observable.
While recent advances in zero-knowledge proofs~\cite{panja2018secure,wang2018designing} and multi-party computations~\cite{raman2018trusted,benhamouda2019supporting} hold many promises, these technologies are yet to gain traction in the context of DeFi.
One of the main friction points is the large computational cost of these technologies, which make them very expensive to use and deploy in the context of DeFi.
A decrease of computational cost of the underlying blockchain will be key to how widely privacy-preserving technologies can be deployed by DeFi protocols.

%% file: content/7_conclusion.tex
\section{Conclusion}
\label{sec:conclusion}

In this SoK we have considered DeFi from two points of view, the DeFi Optimist and the DeFi Pessimist, and examined the workings of DeFi systematically. 
First, we laid out the primitives for DeFi before categorizing DeFi protocols by the type of operation they provide.
% We examined the security challenges protocols are exposed to by making a distinction between technical and economic security risks.
After distinguishing between different types of information relevant to a DeFi protocol, we provided a working definition of an exploit. 
We established economic security on the same level as technical security and created a novel functional categorization of the different types of risk therein. 
Further, we provided clear definitions of these risks as well as insights into the types of models that are needed to understand and defend against these risks.
% In so doing, we were able to systematize attacks that have been proposed in theory and/or occurred in practice into categories of attacks that either rely on an agent's ability to generate risk-free profits by exploiting the technical structure of a blockchain or to game the incentive structure of a protocol to obtain a profit at the expense of the protocol.
Finally, we drew the attention to open research challenges that require a holistic understanding of both the technical and economic risks.

While DeFi may have the potential to creating a permissionless and non-custodial financial system, the opinion put forward by the DeFi optimist, the open technical and economic security challenges remain strong.
The DeFi pessimist is, at least for now, on firm ground: solving these challenges in a robust and scalable way is a central challenge for researchers and DeFi practitioners.
In the end, however, it is the blend between promise and challenge --- the tension between the views of the DeFi optimist and the DeFi pessimist --- that makes DeFi a worthwhile and exciting area for research.

%% file: content/appendix.tex
\clearpage
\appendix

\section{DeFi Protocols}
\label{appendix:defi-protocols}

% \appendix
\hidelinks
\begin{table}[hbt]
    \caption{DeFi Protocols: A selection of prominent DeFi protocols classified according to the proposed protocol types.}
    \begin{tabularx}{\columnwidth}{XXXXX}
        \toprule
        Exchanges                        & PLFs &   Stablecoins & Portfolio Managers & Derivatives  \\
        \midrule
        
        Curve & Compound & Maker & Harvest & Opyn\\
        
        Uniswap & Aave & Unit & Yearn & Hegic   \\
        
        Sushiswap & dYdX & Reflexer & Set & Synthetix \\
        
        Balancer & Cream & Fei & Alpha \\
        
        Bancor & & & \\
        
        1inch & & & \\
        
        \bottomrule
    \end{tabularx}
    \label{table:defi-existing-protocols}
\end{table}
\restorelinks

\section{Empirical Exploits}
\label{app:empirical-exploits}
%% Reentrancy
There have been a range of exploits in DeFi applications.
This is a non-exhaustive list of some of the exploits and vulnerabilities referenced in Sections~\ref{sec:technical-security} and~\ref{sec:economic-security}. 

\point{Reentrancy Exploits}
\begin{itemize}
    \item \textit{dForce:} One of the most prominent examples of this exploit was against the dForce protocol~\cite{web:dforce}, which features a PLF, in April $2020$ to drain around $25$ million USD worth of funds~\cite{web:dforce-hack}.The attacker used imBTC~\cite{web:imbtc}, which is an ERC-777 token~\cite{web:eip777}, to perform the attack. A particularity of ERC-777 tokens, as opposed to ERC-20 tokens, is that they have a hook calling the receiver when the receiver receives funds. This means that any ERC-777 tokens will indirectly result in the receiver having control of the execution. In the dForce attack, the attacker used this reentrancy pattern to repeatedly increase their ability to borrow without enough collateral to back up their borrow position, effectively draining the protocol's funds.

\item \textit{imBTC Uniswap Pool:} Despite the fact that Uniswap does not support ERC-777 tokens~\cite{whitepaper:uniswap}, an imBTC Uniswap~\cite{web:uniswap} pool worth roughly $300\,000$ USD was drained using the reentrancy attack.
\end{itemize}

\point{Integer Manipulation Exploits}

\begin{itemize}
    \item \textit{YAM:} In August $2020$, the YAM protocol~\cite{web:yam}, which had locked almost $500$ million USD worth of tokens in a very short period of time, realized that there was an arithmetic-related bug.Two integers scaled to their base unit were multiplied and the result not scaled back, making the result orders of magnitude too large~\cite{web:yam-bug,web:yam-bug-analysis}.This prevented the governance to reach quorum and locked all the funds in the protocol's treasury contract, effectively locking over $750\,000$ USD worth of tokens~\cite{medium:yam_finance} indefinitely.
\end{itemize}

\point{Logical Bug Exploits}

\begin{itemize}
    \item \textit{bZx:} In September $2020$, the bZx protocol~\cite{web:bzx}, a lending protocol, suffered a loss of over $8$ million USD due to a trivial logic error~\cite{web:bzx-attack-analysis}, despite having been through two independent audits. The bZx protocol uses its own ERC-20 tokens, which are minted by locking collateral and repaid to redeem the locked collateral. As other ERC-20 tokens, bZx tokens allow users to transfer the tokens. However, due to a logical bug, when a user transferred tokens to themselves, the amount transferred would effectively only be added to their balance, and not correctly subtract from it, allowing a user to double his amount of tokens at will. The tokens created could then be used to withdraw funds that the attacker never owned or locked.
\end{itemize}

\point{Market Manipulation}
\label{app:empirical-exploits:market-manip}
\begin{itemize}
    % \item \textit{Compound:} To illustrate the potential of market manipulation attacks, the stablecoin DAI, which historically has thin liquidity, traded at a temporary price of \$$1.30$ over a course of about $20$ minutes on Coinbase Pro, a major centralized cryptoasset exchange, before returning to its intended \$$1$ peg~\cite{khatri20dai}. As a result, the Compound Open Price Feed~\cite{web:open-price-feed}, a cryptoasset price oracle which is in part based on prices signed by Coinbase, reported a DAI price of \$$1.23$ to Compound for a short period of time. This incident triggered (arguably wrongful) liquidations on collateral worth approximately \$$89$m, costing the liquidated Compound borrowers $23$\% (from the imbalanced DAI price) plus an additional $5$\% (the Compound liquidation incentive, i.e., the discount at which collateral is sold at during a liquidation) on their liquidated assets. While the incident was not clearly an exploit, market structure could be exploited in this way to profit from the triggered liquidations.
    
    \item \textit{Venus:} Since pre-printing this paper, a clear exploit that manipulated this market structure was performed on the Venus protocol \cite{venus-may21}. In this exploit, the attacker manipulated the thinly traded XVS market and borrowed large amounts of BTC against XVS collateral at the manipulated high prices. This led to \$100m of bad debt (effectively, the profit for the attacker) in the protocol when the XVS market equilibrated to normal pricing.
\end{itemize}

\section{Batch Settlement Systems}
\label{app:batch-settlement-gnosis}
In Gnosis exchange~\cite{gnosis}, trades are matched algorithmically in periodic batches maintained by decentralized keepers. Keepers compete to solve a complicated matching problem. They submit solutions on-chain, from which the protocol executes the best solution, by some metric. If this keeper market is competitive, trades should be settled at fair prices, though issues can arise when the keeper market is not competitive \cite{gnosis_mta_issue} or if the method for choosing the best keeper solution can be gamed \cite{gnosis_api3}.